\documentclass[11pt]{article}   
\usepackage[T1]{fontenc}
\usepackage{physics}
\usepackage{graphicx}
\usepackage[margin=1in]{geometry}      
\usepackage{caption}
\usepackage[colorlinks]{hyperref}
\usepackage{doi}
\usepackage[nottoc,notlot,notlof]{tocbibind}
\usepackage[sort&compress,numbers]{natbib}
\usepackage{amssymb,amsmath,graphicx,color,amsfonts}
\usepackage[tight]{subfigure}		
\usepackage{amssymb, authblk}

\hypersetup{
	bookmarksnumbered,
	pdfstartview={FitH},
	citecolor={darkgreen},
	linkcolor={darkblue},
	urlcolor={darkblue},
	pdfpagemode={UseOutlines}}
\definecolor{darkgreen}{RGB}{20,100,20}
\definecolor{darkblue}{RGB}{0,0,130}
\definecolor{darkred}{rgb}{.8,0,0}

\title{High harmonic spectroscopy of quantum phase transitions in a high-T$_c$ superconductor}

\author[1]{Jordi Alcal\`{a}}
\author[2,$\dagger$]{Utso Bhattacharya}
\author[2,3,*]{Jens Biegert}
\author[4,5]{Marcelo Ciappina}
\author[2,$\dagger$]{Ugaitz Elu}
\author[2]{Tobias Gra\ss}
\author[2,6,7,8]{Piotr~T. Grochowski}
\author[2,3]{Maciej Lewenstein}
\author[1]{Anna Palau}
\author[2]{Themistoklis P. H. Sidiropoulos}
\author[2]{Tobias Steinle}
\author[2]{Igor Tyulnev}

\affil[1]{ICMAB-CSIC - Institut de Ciència de Materials de Barcelona, Campus UAB, 08193 Bellaterra, Spain}
\affil[2]{ICFO - Institut de Ciencies Fotoniques, The Barcelona Institute of Science and Technology, 08860 Castelldefels (Barcelona), Spain}
\affil[3]{ICREA, Pg.~Lluís Companys 23, 08010 Barcelona, Spain}
\affil[4]{Guangdong Technion - Israel Institute of Technology, Shantou, 515063 Guangdong, China}
\affil[5]{Technion - Israel Institute of Technology,  32000 Haifa, Israel}
\affil[6]{Center for Theoretical Physics, Polish Academy of Sciences, Aleja Lotnik\'ow 32/46, 02-668 Warsaw, Poland}
\affil[7]{Institute for Quantum Optics and Quantum Information, Austrian Academy of Sciences, A-6020 Innsbruck, Austria}
\affil[8]{Institute for Theoretical Physics, University of Innsbruck, A-6020 Innsbruck, Austria}
\affil[$\dagger$]{These authors contributed equally}
\affil[*]{Corresponding author: jens.biegert@icfo.eu}

\begin{document}

\maketitle

\abstract{We report on the new non--linear optical signatures of quantum phase transitions in the high-temperature superconductor YBCO, observed through high harmonic generation. While the linear optical response of the material is largely unchanged when cooling across the phase transitions, the nonlinear optical response sensitively imprints two critical points, one at the critical temperature of the cuprate with the exponential growth of the surface harmonic yield in the superconducting phase, and another critical point, which marks the transition from strange metal to pseudogap phase. To reveal the underlying microscopic quantum dynamics, a novel strong-field quasi-Hubbard model was developed, which describes the measured optical response dependent on the formation of Cooper pairs. Further, the new theory provides insight into the carrier scattering dynamics and allows to differentiate between the superconducting, pseudogap, and strange metal phases. The direct connection between non--linear optical response and microscopic dynamics provides a powerful new methodology to study quantum phase transitions in correlated materials. Further implications are light-wave control over intricate quantum phases, light-matter hybrids, and application for optical quantum computing.}

\section{Introduction}\label{sec1}

Attosecond technology~\cite{Krausz2009}, specifically the process of high harmonic generation (HHG)~\cite{McPherson1987,Ferray1988,Corkum1993}, provides an all-optical probe of the microscopic dynamics of atoms, molecules and solids. Shortly after the first observation of high harmonics in atoms, their generation was understood~\cite{Kulander1993,Maciej1994,Corkum1993} as arising from electron recollision after strong field photo-ionisation and excursion in the continuum. Since the harmonic signal strongly depends on the electron recollision angle and time, high-harmonic spectroscopy(HHS) is a sensitive nonlinear probe of microscopic electronic structure with atomic spatial and sub-optical cycle temporal resolution. HHS of solids~\cite{Ghimire2011, Hohenleutner2015}, 2D materials~\cite{Yoshikawa2017,Baudisch2018}, or nano-structured media~\cite{Vampa2017,Franz2019} differs from the gas phase since the optical-field-driven electronic wavepacket is de-localised over many lattice sites, the wavefunction depends on the lattice momentum, and a hole has to match the electron's momentum for recombination to occur~\cite{Lanin2017,Luu2018}. Recent experimental efforts extended HHS as non-perturbative probe to quantum materials~\cite{Yoshikawa2017,Baudisch2018,Silva2019} and to topological insulators~\cite{Maczewsky2020,Schmid2021,Baykusheva2021}.

The sensitivity of HHS to the intricate microscopic details of carriers and lattice predestines HHS to investigate strong interactions and quantum correlations which lead to fascinating new states of matter such as superconductivity. The phase transition into a strongly-correlated superconductive state is described by the spontaneous symmetry breaking of the U(1) redundancy when cooling below the critical temperature $T_\text{c}$ of the material. As we will show, HHS is a novel sensitive probe of the dynamic evolution of the superconducting phase transition since the formation of composite bosons by pairing two fermionic spin-1/2 particles (Cooper pairs) changes the distribution of charge carriers and this sensitively registers in the high harmonic amplitudes and spectral distribution. Moreover, we will show that HHS can identify additional phase transitions between quantum phases in the strongly correlated material which are not accessible through the linear optical response and they are difficult to detect with established methods such as SQUID magnetometry or four-probe transport measurements. 

A conventional superconductor can be described by the Anderson-Higgs mechanism, which explains that an optical nonlinear response is due to a gapless phase mode (Nambu-Goldstein) and a gapped amplitude mode (Higgs) of the ordering parameter. In the simplest case, and depending on the strength and type of excitation, Boltzmann and Ginzburg-Landau theories~\cite{Stephen1965,Ginzburg1965} predict a second-order response, which mixes with the excitation mode~\cite{Barankov2006,Yuzbashyan2006}, thus the generation of the third harmonic~\cite{Seibold2021}. Unconventional high-$T_\text{c}$ superconductors are of tremendous interest for a wide range of applications ranging from electronic devices and information processing devices to optical quantum computers and quantum simulators. However, due to their rich landscape of quantum phases and the difficulties of experimental methods to probe the microscopic dynamics, our understanding is still very limited. Therefore, the development of all-optical and ultrafast probes of the macroscopic dynamics inside such materials, which is compatible with existing methods, is highly desirable. To this aim, we apply HHS to investigate the transition between the different phases of the unconventional high-$T_\text{c}$ superconductor YBa$_2$Cu$_3$O$_{7-d}$ (YBCO). We elucidate the connection between the measured optical spectra, the transition between strange metal and pseudogap phases and the superconducting phase transition with a novel strong-field Hubbard model. The HHS measurement clearly shows a departure from the normal conducting phase with an increased formation of Cooper pairs upon cooling. The variation in harmonic orders is linked to phenomenological energy and phase relaxation times, which identify the transition to the fluctuation regime~\cite{Coton2010,SolovEv2009}, i.e. between the strange metal and pseudogap phases, and the sudden transition at $T_\text{c}$ into the superconducting phase. Unconventional superconductors, like YBCO, are material systems in which the formation of composite bosons out of paired fermions is not mediated by phonon exchange, but by some other kind of energy exchange~\cite{Sigrist1991}, for instance, due to spin fluctuations. Such systems present many standing fascinating questions. It is thus important to have new powerful experimental techniques like HHS that provide a fresh and alternate view of the problem. 

\section{Results and Discussion}\label{sec2}

\begin{figure}[h]
\captionsetup{width=0.9\textwidth}
\centering
\includegraphics[width=0.9\textwidth]{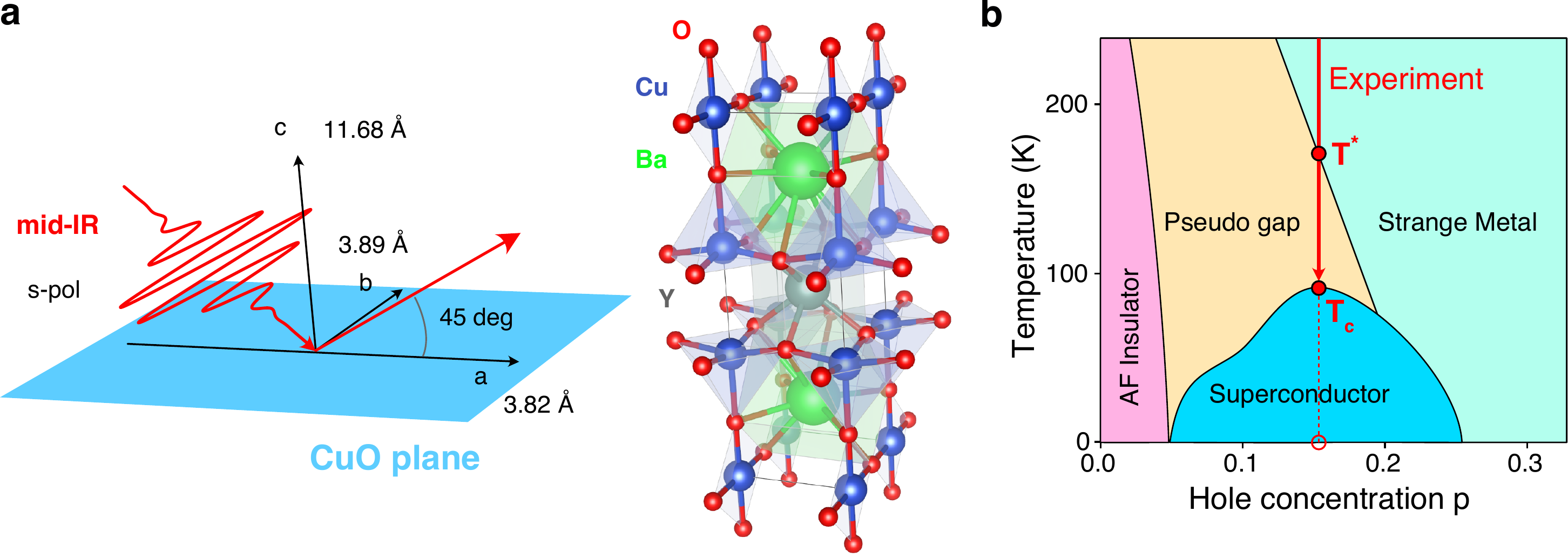}
\caption{\textbf{Experimental setup and YBCO properties:} (a) Orthorhombic unit cell of YBa$_2$Cu$_3$O$_7$ (YBCO) together with the crystallographic axes. YBCO consists of alternating planes along the crystallographic c-axis with order CuO--BaO--CuO$_2$--Y--CuO$_2$--BaO. Most relevant for superconductivity are the copper-oxide planes. Also indicated is the experimental arrangement with linearly-polarised mid-IR laser fields whose electric field vector lies along the copper-oxide planes. (b) shows the phase diagram for YBCO. Based on the value of $T_\text{c}$ from the measurement, we determine a hole concentration of p $\sim$ 0.15. Based on this value, we mark the transition between the different phases by the red arrow. In accord with measurement and theory, $T^\text{*}$ marks the transition between strange metal and pseudogap phases at 173 K.}\label{fig1}
\end{figure}

YBa$_2$Cu$_3$O$_{7-d}$ (YBCO) is a copper-oxide cuprate that crystallises in an orthorhombic perovskite structure. The YBCO unit cell consists of two CuO$_{2}$ planes which sandwich an oxygen-depleted CuO$_{d}$ plane; see Fig.~\ref{fig1}(a). The CuO$_2$ planes are central to superconductivity, interacting with oxygen-depleted CuO$_{d}$ chains which act as charge reservoirs~\cite{Gauquelin2014,Saxena2012,Hirao1992}, and specifically the interaction of electrons in the $2p_x$ and $2p_y$ orbitals of oxygen (O) ions and the $3d_{x^2-y^2}$ orbitals of copper (Cu) ions. Thus, YBCO is a quasi-two-dimensional superconducting system. As a result of the strong Coulomb repulsion by the Cu$^{3+}$ ions, the $3d_{x^2-y^2}$ band splits into lower (LHB) and upper Hubbard bands (UHB) with a gap of up to 2 eV, and the density of states exhibits a peak at the top of the LHB. The transition to the SC phase is presently understood as hole--doping the oxygen $2p$ sites altering the exchange interaction between the copper spins which mediates the formation of Cooper pairs. Our measurement reveals two critical temperatures (discussed below) which allow determining the doping level of the material at p $\sim$ 0.15. Figure~\ref{fig1}(b) depicts the phase diagram of YBCO and, based on the determined doping level for our YBCO sample, the red arrow indicates that our measurement transitions between the strange metal phase at room temperature into the pseudogap phase at $T^\text{*}$ and into the SC phase at $T_\text{c}$.

For our investigation, we measure high harmonics in reflection from 100--nm--thick films of YBCO which are mounted on a Joule-Thomson micro--refrigerator. The superconducting properties of the YBCO films were characterised by inductive measurements and a critical temperature of 88 K was obtained, indicating proper oxygen doping; see Fig.~\ref{figSI2} and description of the SI.
Having confirmed the high quality of the YBCO films and the transition to the SC phase, we induce non-perturbative high harmonic generation with sub-bandgap, ultrashort mid-infrared (mid-IR) pulses at 3200 nm (0.4 eV) from a mid-IR OPCPA~\cite{Ugaitz2017,Elu2020}. The linearly-polarised 94--fs, mid-IR pulses are focused to a vacuum electric field strength up to 0.083 V/{\AA} and they intersect the material at 45 deg under s-polarisation. Due to the chosen reflection geometry with s-polarisation we avoid the generation of even order surface harmonics, arising due to symmetry breaking at the interface, and detect the odd-order harmonics (HH3, HH5 and HH7) of the fundamental photon energy at 0.4 eV. This measurement geometry ensures the best possible signal-to-noise ratio as the fundamental's energy is exclusively distributed amongst odd harmonics orders. To prevent oxidation of the samples, and unwanted alteration of the superconductive properties, we conduct all experiments inside a vacuum chamber. We have checked that the 400--$\mu$m--thick CaF$_{2}$ windows of the vacuum chamber do not contribute to the measured signal. The YBCO sample is cooled with the Joule Thomson refrigerator to a minimum temperature of 77 K and we record the reflected radiation by imaging onto the entrance slit of a spectrograph for analysis.

\begin{figure}[h]
\captionsetup{width=0.9\textwidth}
\centering
\includegraphics[width=0.9\textwidth]{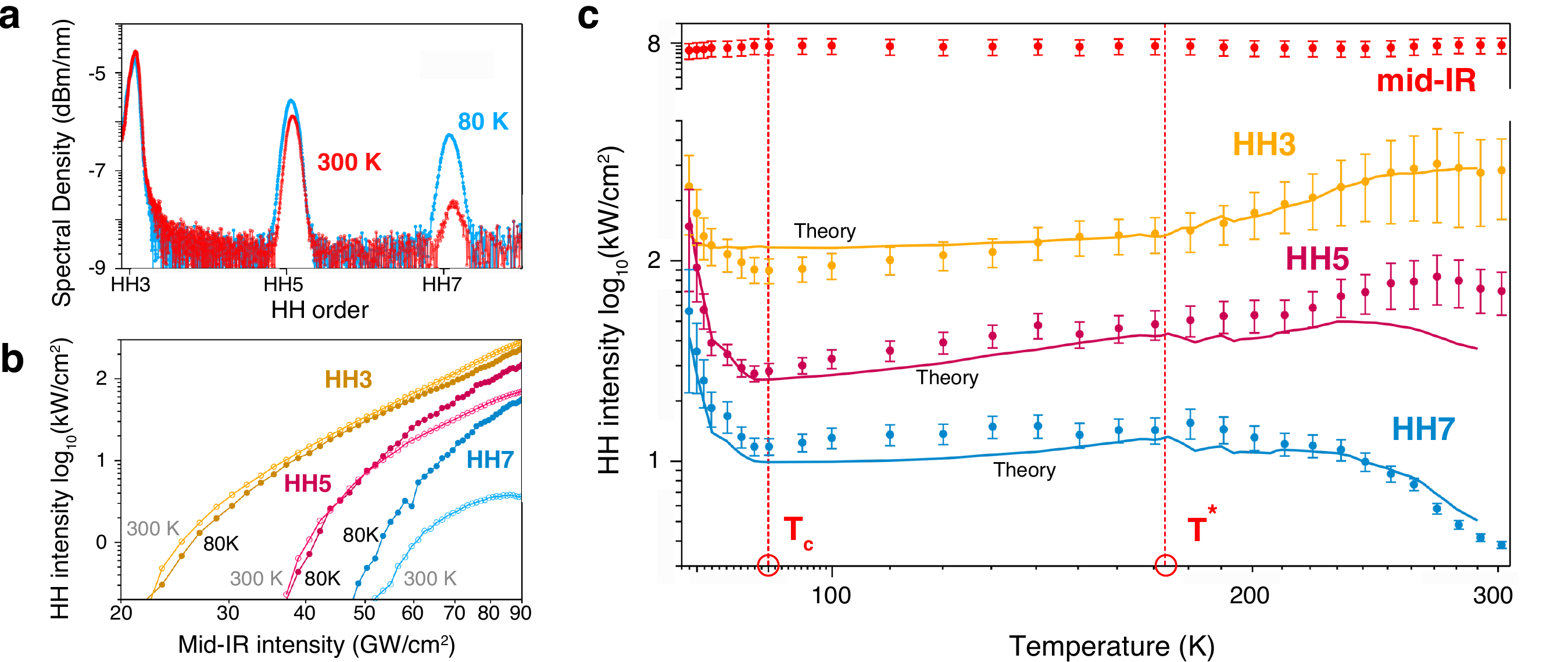}
\caption{\textbf{High harmonic spectroscopy of YBCO:} (a) Harmonic spectra showing odd orders HH3, HH5 and HH7 for room temperature (red) and at $T_\text{c} = 90 K$ (blue) for a mid-IR field strength of 0.083 V/{\AA}. Visible is already a blueshift of room-temperature harmonics with increasing harmonic order and relative to the harmonics measured at $T_\text{c}$. We observe a dramatic increase of HH7 amplitude upon cooling into the SC phase. (b) Shown is the scaling of harmonic order with mid-IR peak-intensity for measurements at room temperature and at $T_\text{c}$. Dark colours show data for the SC phase whereas light colours indicate data at room temperature. (c) Shown is the reflected mid-IR field together with harmonics HH3, HH5 and HH7 as a function of temperature. These measurements are taken for a mid-IR field strength of 0.083 V/{\AA}. Results from the strong-field quasi-Hubbard model are overlaid as solid lines. We observe a dramatic increase of HH7 amplitude upon cooling into the SC phase at 88 K. All harmonic orders show a clear turning point at the critical temperature $T_\text{c}$ and an exponential increase in amplitude. More subtle, but still clearly discernible is another critical point $T^\text{*}$ at 173 K, marking the transition from strange metal to the pseudogap phase. The functional behaviour is reproduced by the model.}\label{fig2}
\end{figure}

Figure~\ref{fig2}(a) shows measured harmonic spectra at room temperature (red curve) and close to the critical temperature $T_\text{c}$ (blue curve). As expected, we record only odd-order harmonics. Clearly noticeable is a blueshift of room-temperature harmonics with increasing harmonic order and relative to the harmonics measured at $T_\text{c}$. Next, we record how HH3 (yellow), HH5 (red), and HH7 (blue) scale as a function of the mid-IR peak intensity; shown in Fig.~\ref{fig2}(b).

While it is expected that all harmonics increase for a higher peak intensity of the fundamental, we instead observe a surprising and dramatic increase for HH7 for the highest mid-IR peak intensity at 90 GW/cm$^{2}$. We thus fix the mid-IR peak intensity at 90 GW/cm$^{2}$, corresponding to a vacuum electric peak field amplitude of 0.083 V/{\AA}, and record the optical response of the material as a function of temperature. 

Figure~\ref{fig2}(c) shows the results of this measurement. While we find that the fundamental (red dots) does not vary significantly with temperature, all the harmonic orders exhibit varying and strong trends with temperature. Most strikingly, HH7 increases in amplitude by over one order of magnitude over the measured temperature range while HH3 and HH5 amplitudes initially decrease with temperature. For instance, the amplitude of HH3 decreases until $T^\text{*}$ = 173 K after which the amplitude varies very little. A similar trend is observed for HH5. In stark contrast, HH7 increases by at least a factor of 5 until reaching the critical point. This critical point marks the transition between strange metal into the pseudogap phase, at $T^\text{*}$ = 173 K, and the amplitude of all harmonics varies very little. Further cooling to $T_\text{c}$ = 88 K, a dramatic exponential increase is measured for all harmonic orders. This marks the quantum phase transition into the superconducting state. It is interesting to note that the highest nonlinear measure, HH7, is the most sensitive indication of all critical points to investigate the quantum phase transitions from strange metal into pseudogap and SC phases. We note that measurements of the magnetisation of the sample readily confirm the temperature value at which the SC phase transition occurs; see SI. However, in contrast to HHS, the identification of the transition between strange metal and pseudogap quantum phases is not as obvious from conventional methods like a transport measurement; see Fig.~\ref{figSI2}. 
To further elucidate the connection between measured optical nonlinearity and the formation of Cooper pairs in the superconductor, we have developed a new theoretical approach that describes HHG in a high-$T_\text{c}$ superconductor through a two-band quasi-Hubbard model with BCS d-wave pairing. From the theoretical perspective, high-$T_\text{c}$ superconductivity is often described in terms of the Hubbard model~\cite{Hubbard1963}, a paradigmatic framework in strongly correlated quantum systems capturing both electronic and spin interactions. There have been many theoretical approximations and extensions of the Hubbard model to study the cause of high-$T_\text{c}$ superconductivity~\cite{Emery1987,Zhang1988,Hirsch1987}. In the literature, high-$T_\text{c}$ superconductivity is often modelled with intra- and inter-orbital hopping and electron-electron interactions by a three-band Hubbard model~\cite{Kung2016,Scalettar1991}. Due to numerical challenges, simplified one-band versions have been considered~\cite{Zhang1988}. But, such a simple model does not actively account for the hopping between oxygen orbitals in a non-equilibrium scenario. In contrast, our two-band quasi-Hubbard model uses a band structure calculated from density functional theory (DFT), combined with attractive Hubbard-like interactions in the mean-field limit. The electrons in the lower band interact with each other and can therefore be well described by an attractive Hubbard interaction, giving rise to the Cooper pairing and superconductivity.
Transitions between the higher band and lower band are to describe high-energy electron hopping processes between the oxygen orbitals. We assume that electrons in the higher band are non-interacting and dipolar coupling includes intra- and inter-band transitions. The dynamic evolution of the system is described by semiconductor Bloch equations, including electron-electron and electron-phonon scattering processes through phenomenological dephasing terms; see SI for details on the model.

The calculated high harmonic spectra are shown in Figure~\ref{fig2}(c) (solid lines). We find an excellent match with the experimental data. The model faithfully reproduces the functional form of the measurement data over the entire temperature range and for several orders of magnitude of harmonic amplitude. Most importantly, the model describes the position of critical points, changes of slopes, and the position and trend of exponential harmonic amplitude. The excellent match of simulations with data permits identification of the critical point at $T^\text{*}$ = 173 K as the transition from the strange metal phase into the pseudogap phase, thus marking the beginning of the fluctuation regime. Microscopically, the transition to the pseudogap phase, by lowering the temperature, is accompanied by strong fluctuations associated with a quantum critical point~\cite{Shekhter2013} which is clearly identified by our model and the measurement; see Fig.~\ref{fig2}(c). Lowering the temperature further, we expect a further increase in carrier scattering time and a respective increase of the current. 

\begin{figure}[h]
\captionsetup{width=0.9\textwidth}
\centering
\includegraphics[width=0.9\textwidth]{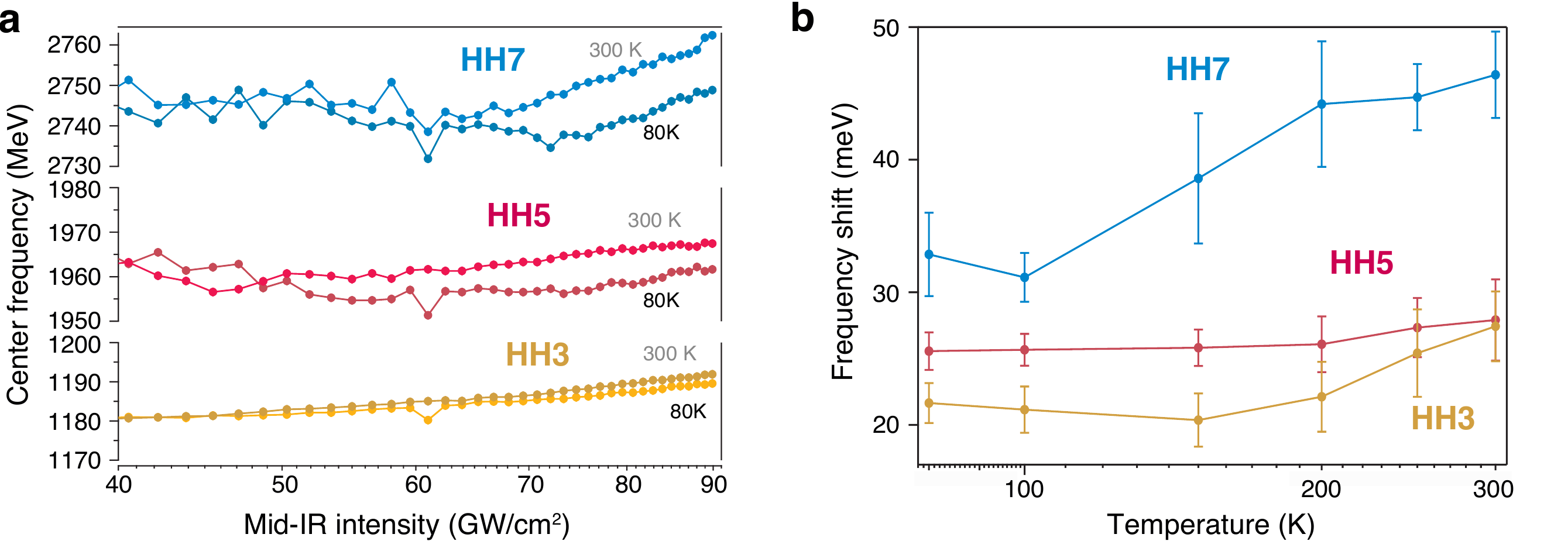}
\caption{\textbf{Harmonic frequency and shift:} (a) 
The center frequencies of the 3rd, 5th, 7th harmonic peak are plotted vs. the intensity of the exciting light at 300 K and at 80 K. A frequency blueshift increasing with the driving intensity is observed. (b) The temperature-dependence of the blueshift is shown. All harmonics show a rise of the blueshift for temperatures $T\gtrsim T^*$ is observed. This behaviour reflects the suppression of scattering processes in the superconducting phase and the pseudogap phase.
\label{fig3}}
\end{figure}

To this end, we extract the blueshift of HHS across all the three regimes: superconducting, pseudogap and strange metal. The blueshift of HHS is reminiscent of strong electron-electron scattering during HHG~\cite{Baudisch2018} and occurs due to a back action of the quasiparticle dynamics on the laser field. The strong carrier scattering and consequential blue--shifting are reminiscent of effective energy loss and dephasing of the HHG process. As such it can reveal crucial details about the quasiparticles in each strongly correlated regime. Figure~\ref{fig3} shows measurement results which show that at room temperature, carrier scattering is strongest and the highest nonlinearity, here HH7, is affected most; also see Fig.~\ref{fig1}(b). Interestingly, the directly measurable blueshift, while clearly discernible in all harmonic orders, does allow to clearly identify the critical temperature and transition between the phases. By comparing Figs.~\ref{fig2} and~\ref{fig3}, within the sensitivity of our measurement, the harmonic amplitudes prove to be the strongest observable for carrier scattering. 

\begin{figure}[h]
\captionsetup{width=0.9\textwidth}
\centering
\includegraphics[width=0.9\textwidth]{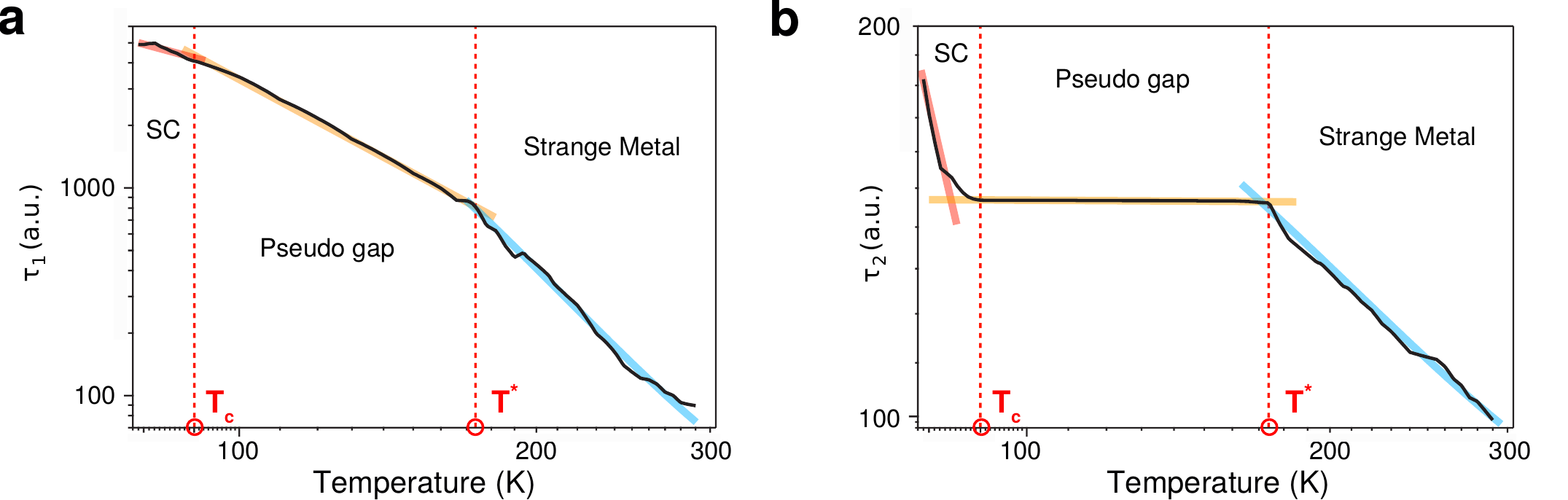}
\caption{\textbf{Phenomenological scattering parameters:}  (a,b) By fitting the experimental data to the model, we extract the temperature-dependence of the phenomenological scattering and dephasing times $\tau_{1}$ and $\tau_{2}$. Clearly visible are changes in the functional behaviour of the two parameters at the two critical points. In the given temperature range from 80 K to 300 K, the scattering time $\tau_1$ decreases from 1400 to 100 a.u. (34 to 2.4 fs), and the dephasing time $\tau_2$  decreases from 180 to 100 a.u. (4.4 to 2.4 fs)
}\label{fig4}
\end{figure}

Our model provides insight into this dynamics with phenomenological scattering and dephasing times $\tau_{1}$ and $\tau_{2}$, shown in Fig.~\ref{fig4}. We note that strong field excitation as in HHS precludes the straightforward interpretation of $\tau_{1}$ and $\tau_{2}$ with inverse resistivity. This would be straightforward in the linear excitation regime, but the association with a resistivity under strong-field excitation requires further theoretical development. Nevertheless, both values $\tau_{1}$ and $\tau_{2}$ allow identifying the quantum critical points. 

Lowering the temperature to 88 K, a striking change of high harmonic amplitude and an exponential increase is observed. This changeover clearly marks the quantum phase transition into the superconducting phase. The theoretical model reproduces the exponential change of HH7 and HH5 in addition to the overall functional form of the experimental curve and it identifies the dramatic increase in harmonic amplitude due to a vastly increased current from inter-band carrier motion at near-zero resistivity. 
We note that a detailed analysis based on an extended microscopic model that could relate the self-energy, due to scattering processes such as electron-electron, electron-phonon, etc., with the measured blueshift would provide a more quantitative picture of the transition from a pseudogap to a non-Fermi liquid phase, i.e., the strange metal phase of the cuprates. The quasi-particle weight extracted from the real part of the self-energy could be used to understand whether a system is in a Fermi liquid state or not. This distinction should be possible because, in a Fermi liquid state, the quasi-particle weight is expected to have a finite value at zero temperature, whereas, it is zero for a non-Fermi liquid at zero temperature. A large reduction of the quasi-particle weight inferred from the blueshift would reveal the physics of the transition towards the non-Fermi liquid strange metallic phase.

\section{Conclusion}\label{sec13}

Our experimental and theoretical findings provide a first striking example of how HHS can be used to distinguish strongly correlated phases of matter and pave the way towards a refined understanding of the intriguing physical processes at work in a high-$T_c$ superconductor. HHS is a technique that is simple to apply but needs a theoretical model to provide microscopic insight. We have provided such a new strong-field quasi-Hubbard model which connects the optical spectrum as observable with the microscopic dynamics of carriers. The model describes the formation of composite bosons from fermion spin-1/2 particles by Cooper pairing and the electron hopping processes between orbitals of the material. The combination between experiment and theory permits extracting phenomenological scattering times which clearly identify the quantum critical points. We note that in the linear response regime, these phenomenological amplitudes and phase decay times are directly related to the material's conductivity. Our BCS-type model faithfully models the measured HHS which are shown to be sensitive to all quantum critical points. Our investigation provides the first strong-field investigation of a high $T_\text{c}$ superconductor and measurement of its quantum phase transitions. The method clearly shows the efficacy to identify the different quantum phases, such as the pseudogap phase, the strange metal and the superconducting phase, together with their quantum critical points. We believe that attosecond technology applied to quantum materials provides a novel view of the multi-body physics of strongly correlated materials and it allows to probe the pairing glue of Cooper pairs. Such insight will further our understanding of the strong-field and light-field control of light-matter hybrids with the distinct possibility to control competing charge orders in quantum matter. We envision leveraging strong field control to enable or inhibit superconductivity and to switch quantum phases at the frequency of light. This has implications to develop new ways for energy-efficient information processing and optical quantum computation.

\section{Supplementary Information}\label{SI}

\subsection{Theoretical model}

The theoretical description is based on a Hubbard-like two-band model in the BCS limit, whose time-dependent Hamiltonian reads
\begin{eqnarray}
H(t) & = & \sum_{k,\lambda,\sigma}\varepsilon_{k\lambda}c_{k\sigma\lambda}^\dagger(t)
c_{k\sigma\lambda}(t) + \sum_k \lbrack \Delta_k(t) c_{k\uparrow \rm L}^\dagger (t) c_{-k\downarrow \rm L}^\dagger(t) + {\rm h.c.}\rbrack  \nonumber\\
&~& - E(t)\sum_{k,\lambda,\lambda',\sigma} \lbrack c_{k\sigma \lambda}^\dagger(t)D_{\lambda\lambda^{'}}(k)c_{k\sigma \lambda'}(t)\rbrack.
\label{ham}
\end{eqnarray}
Here, $\lambda \in \{\rm L,U\}$ is the band index for the lower and upper band, $\sigma\in \{\downarrow,\uparrow \}$ is the spin index, and $k$ is the index for pseudomomentum. The operators $c_{k\sigma\lambda}(t)$  and $c_{k\sigma\lambda}^{\dagger}(t)$ are the corresponding time-dependent annihilation and creation operators in the Heisenberg frame. The band energies are given by
$\varepsilon_{k\rm U} = E_{g} + E_{\rm U}\left( k \right) - \mu$ and $\varepsilon_{k \rm L} = E_{\rm L}\left( k \right) - \mu$, where $\mu$ is the chemical potential, determined self-consistently.
The renormalized band structure along the main symmetry direction is obtained from DFT, and parametrized by the functions $E_{\lambda}\left( k \right)$, and a direct energy gap $E_{g} = 0.0317$ a.u. at the band edge.

The second term in Eq.~(\ref{ham}) is the pairing term, proportional to the time-dependent superconducting gap $\Delta_{k}(t)$. This term is obtained from an attractive d-wave interaction $U_k= U\lbrack\cos\left( k_{x}a \right) - \cos\left( k_{y}a \right)\rbrack$ of strength $U$, and the gap parameter must fulfill the BCS self-consistency condition $\Delta_k(t) = - U_k\left\lbrack \sum_{k'}{U_{k'}\langle c_{-k' \downarrow \rm L}(t)c_{k' \uparrow \rm L}(t)\rangle} \right\rbrack$.

The last term in Eq.~(\ref{ham}) describes the coupling to the light field, with an electric field pulse 
$E(t) = E_0\sin^{2}{(\omega_0 t/2n_{\text{cyc}}})\sin{(\omega_0 t)}$  of amplitude $E_{0} = 0.004$ in atomic units (a.u.), center frequency $\omega_0 = 0.01425$ a.u., $n_{\text{cyc}} = 8$ cycles. These values are compatible with the experimental estimated parameters of the laser pulse.
The optical coupling is proportional to the covariant derivative~\cite{Ventura2017}, $- iD_{\lambda\lambda'}(k) = \left\lbrack \delta_{\lambda\lambda^{'}}\partial_{k} - id_{\lambda\lambda'(k) }\right\rbrack$,
with $d_{\lambda\lambda'}(k)$
describing the interband dipole moment, or, for
$\lambda = \lambda'$, the Berry connection of the respective band,
which is zero in our system. For the dipole moment we have used $d_{\rm UL}$=14 a.u. in our numerical evaluations.

For linearly polarized driving fields, typically one spatial dimension in the k-space is enough to model the electron-hole dynamics~\cite{Vampa2014}. 
As such, we restrict our numerical simulations to only a single one-dimensional slice of the full k-space. From Eq.~(\ref{ham}), we then derive the Heisenberg equations of motion which describe the dynamics of the relevant correlators, namely populations
$n_{k\lambda\sigma}( t ) = \langle c_{k\sigma\lambda}^{\dagger}( t )c_{k\sigma\lambda}( t ) \rangle$,
normal interband polarizations
$P_{k\downarrow}^{*}( t) = \langle c_{k \downarrow \rm U}^{\dagger}(t)c_{k \downarrow \rm L}(t) \rangle$
and
$P_{k \uparrow}^{*}(t) = \langle c_{k \uparrow \rm U}^{\dagger}(t)c_{k \uparrow \rm L}(t) \rangle$,
as well as the anomalous mixed-spin polarizations
$A_{k \uparrow \downarrow}(t) = \langle c_{k \uparrow \rm U}^{\dagger}(t)c_{- k \downarrow \rm L}^{\dagger}(t) \rangle$
and
$A_{k \downarrow \uparrow}(t) = \langle c_{k \downarrow \rm U}^{\dagger}(t)c_{- k \uparrow \rm L}^{\dagger}(t) \rangle$,
and the superconducting correlations
$S_{k\rm U}(t) = \langle c_{k \uparrow \rm U}^{\dagger}(t)c_{- k \downarrow \rm U}^{\dagger}(t)\rangle$
and
$S_{k\rm L}(t) = \langle c_{k \uparrow \rm L}^{\dagger}(t)c_{- k \downarrow \rm L}^{\dagger}(t)\rangle$.
These ``superconductor Bloch equations'' form a closed set of first-order differential equations.
In order to account also for effects due to electron-electron and electron-phonon interactions and disorder, we include in these equations non-unitary scattering processes with phenomenological scattering times $\tau_1$ and $\tau_2$:
\begin{eqnarray}
i \partial_{t} \langle O(t) \rangle = - \langle \left\lbrack H(t),O(t) \right\rbrack \rangle  - i \frac{ \langle O(t)\rangle }{\tau_{1/2}}.
\end{eqnarray}
$\tau_1$ is used when $O(t)$ represents a population, while $\tau_2$ when $O(t)$ describes a correlation. The scattering times $\tau_1$ and $\tau_2$ and their temperature-dependence are obtained by fitting theoretical results to the experimental data, see main text Fig.~\ref{fig4}.

The initial conditions of the equations of motion are given by the thermal equilibrium when the electric field is off. The band populations are given by the Fermi-Dirac distribution, assuming a half-filled L-band. All polarizations and the superconducting correlator in the U-band are initially zero. The superconducting correlator in the L-band takes nonzero values up to the critical temperature.
Its value is determined self-consistently by solving the model through the Bogoliubov-de Gennes transformation.
With the phenomenologically motivated choice of $U=0.718$ and a half-filled L-band, the model produces the right critical temperature $T_\text{c}=88$ K.

We then time-evolve the system in the optical field, and evaluate the generated electrical current $J$ in the crystal~\cite{Ventura2017}, given by $J(t) = J_{\text{intra}}(t) + J_{\text{inter}}(t)$ with $J_{\text{intra}} =  \sum_{k,\lambda,\sigma}^{}{\partial_{k}\varepsilon_{k\lambda}\langle n_{k\lambda\sigma}\rangle}$
and
$J_{\text{inter}} =  \sum_{k\sigma}2\ d_{\text{UL}}\left( \varepsilon_{k\rm U} - \varepsilon_{k \rm L} \right)\mathrm{Im}(P_{k\sigma})$, assuming a constant interband dipole moment $d_{\text{UL}}$.
Here, we have already assumed a one-dimensional geometry.
From the time-dependent current $J(t)$ we obtain the harmonic spectrum by Fourier transforming it into the frequency domain, with the results shown in Fig.~\ref{fig2}(c) of the main text.

\subsubsection{YBCO band structure}
\begin{figure}[h]
\captionsetup{width=0.9\textwidth}
\centering
\includegraphics[width=0.4\textwidth]{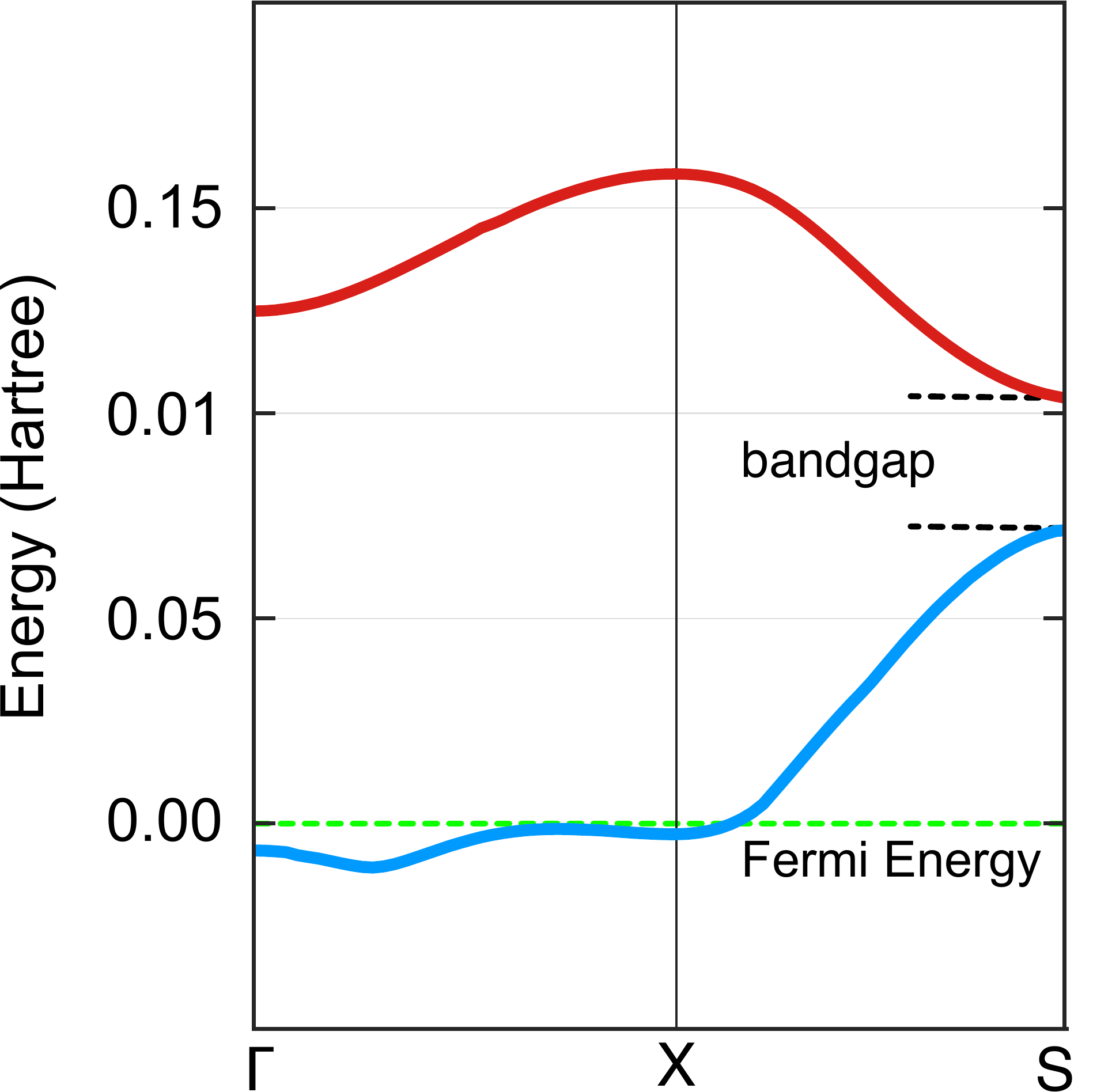}
\caption{\textbf{Band structure:} The two bands $\rm U$ and $\rm L$ are plotted along $\Gamma$--X--S. The horizontal line marks the Fermi energy for a half-filled L-Band.}
\label{figbands}
\end{figure}

The band structure shown in Fig.~\ref{figbands} has been calculated with the all-electron full-potential linearised augmented-plane wave (LAPW) code ELK~\cite{elk}. For simplicity, the $\Gamma$--X--S direction is used for the simulations.


\subsection{High Harmonic Spectroscopy and Data Analysis}
The third (1065 nm), fifth (639 nm) and seventh (457 nm) harmonics generated driven by the 3.2 $\mu$m pulses are measured with a custom configured Maya 2000 Pro high-sensitivity spectrometer from Ocean Insight, Inc. The generated harmonics at the surface of the YBCO sample are imaged in free space onto a 600--$\mu$m--wide spectrometer slit. The spectrum of all three harmonics is measured simultaneous using an integration time of 500 ms. We used a FGB37 bandpass filter from Thorlabs, Inc to measure all the signals together. The purpose of the FGB37 bandpass filter is a calibrated attenuation of the third harmonic signal by $\sim 4$ orders of magnitude such that HH3 can be measured together with the much weaker HH5 and HH7. The filter avoids saturation and possible cross-talk between the array pixels and readout. The high dynamic range of the Maya 2000 Pro spectrometer enables measuring HH5 and HH7 signals without requiring any additional attenuation of HH5. We have conducted additional measurements with a monochromator and calibrated photomultiplier tube to confirm the accuracy of our measurements. The spectrometer's integration time was set to 500 ms to minimise thermal effects during measurement. In addition, an automatic shutter blocked the mid-IR beam for a few minutes between each measurement to ensure thermalisation of the sample to the set temperature before another acquisition was taken. The YBCO temperature was varied from cryogenic to room temperatures and back to avoid systematic errors due to possible hysteresis. This process was repeated numerous times. The standard deviation is calculated from those repeated measurements.

\subsection{YBCO sample fabrication and characterisation}

\begin{figure}[h]
\captionsetup{width=0.9\textwidth}
\centering
\includegraphics[width=0.9\textwidth]{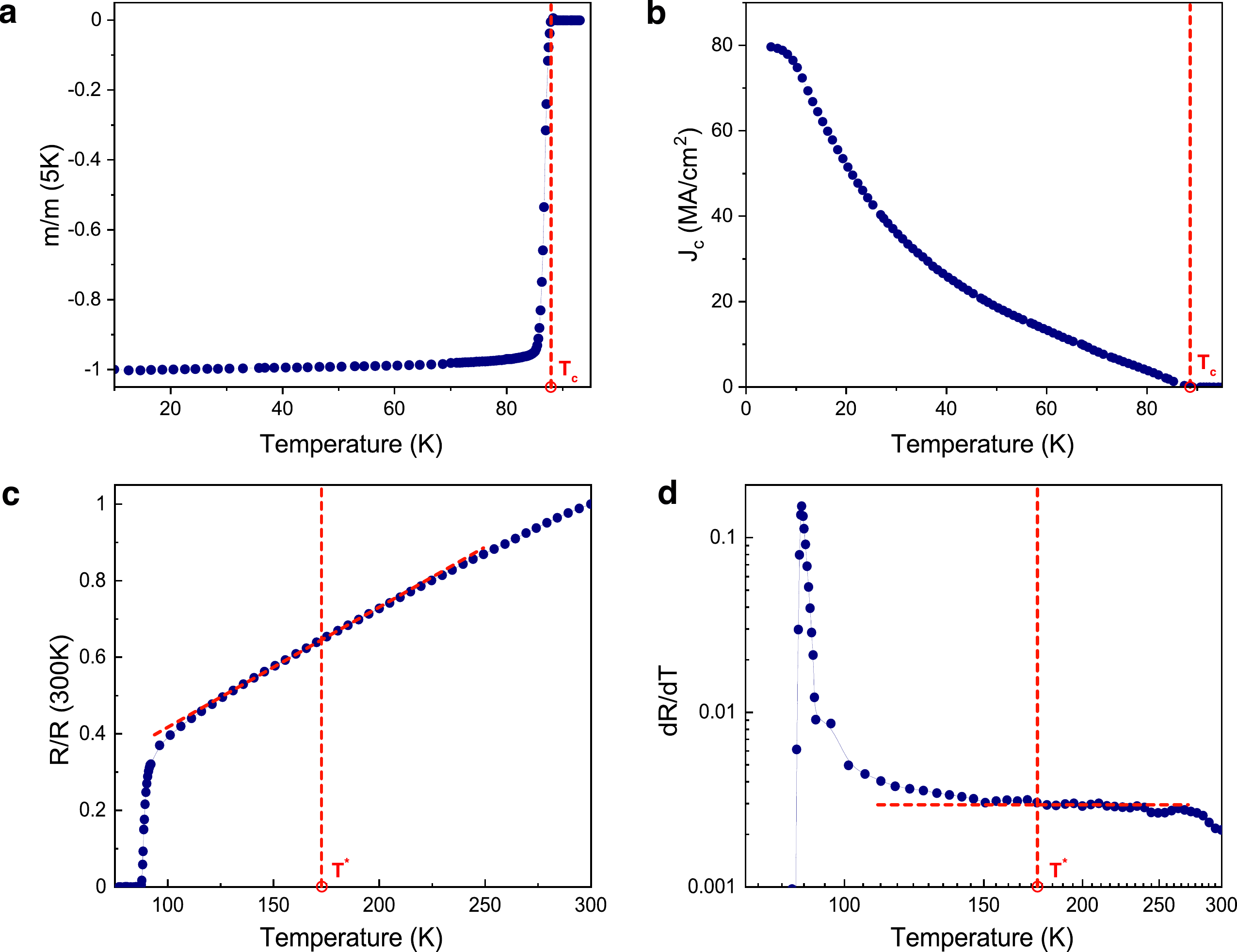}
\caption{\textbf{Characterization of YBCO samples:}  (a) SQUID measurement of the temperature dependence of the magnetic moment, normalised at 5 K, showing a critical temperature of$T_\text{c}$ = 88 K and (b) critical current density as a function of temperature, $J_\text{c}$(T). (c) Shows the resistivity, normalised at 300 K, as function of temperature and (d) shows the derivative. For completeness the HHS determined critical temperatures are superimposed. Solid lines in (c) and (d) are linear fits performed above T$^{*}$.}\label{figSI2}
\end{figure}

YBa$_2$Cu$_3$O$_{7-d}$ (YBCO) films with a thickness of 100 nm were grown using pulsed laser deposition (PLD) on 5×5 mm$^2$ LaAlO$_3$ (100) single crystal substrates with thicknesses of 500 $\mu$m. The films where deposited at $800^\circ$C and PO$_2$ = 0.3 mbar with a pulse frequency of 5 Hz and oxygenated at $600^\circ$C at 1 bar. After film growth the substrate was milled to a final thickness of 50 $\mu$m. SQUID magnetometry (Quantum Design) was used to determine the critical current temperature and critical current density using the Bean critical state model. Transport measurements were performed with four-probe configuration by a commercial physical property measurement systems (PPMS). The results from these measurements are shown in Fig.~\ref{figSI2}. We note that these measurements are in very good agreement with literature values; see e.g. Ref.~\cite{Taillefer2010}.

\section{Acknowledgments}
We thank V. Kunets from MMR Technologies and X. Menino at ICFO for technical support and Dr. K. Dewhurst and Dr. S. Sharma for help with the ELK code.

\subsection*{Funding}

J.B., U.E., T.P.H.S., T.S. and I.T. acknowledge financial support from the European Research Council for ERC Advanced Grant “TRANSFORMER” (788218) and ERC Proof of Concept Grant “miniX” (840010). J.B. and group acknowledge support from FET-OPEN “PETACom” (829153), FET-OPEN “OPTOlogic” (899794), EIC-2021-PATHFINDEROPEN "TwistedNano" (101046424), Laserlab-Europe (654148), Marie Sklodowska-Curie ITN “smart-X” (860553), Plan Nacional PID-PID2020-112664GB-I00-210901; AGAUR for 2017 SGR 1639, “Severo Ochoa” (SEV- 2015-0522), Fundació Cellex Barcelona, the CERCA Programme / Generalitat de Catalunya, and the Alexander von Humboldt Foundation for the Friedrich Wilhelm Bessel Prize. 
U.B., T.G., P.T.G., and M.L. acknowledge funding from the European Research Council for ERC Advanced Grant NOQIA; Agencia Estatal de
Investigación (R$\&$D project CEX2019-000910-S, funded by MCIN/
AEI/10.13039/501100011033, Plan National FIDEUA PID2019-106901GB-I00, FPI,
QUANTERA MAQS PCI2019-111828-2, Proyectos de I+D+I “Retos Colaboración”
RTC2019-007196-7); Fundació Cellex; Fundació Mir-Puig; Generalitat de Catalunya through the CERCA program, AGAUR Grant No. 2017 SGR 134, QuantumCAT  U16-011424, co-funded by ERDF Operational Program of Catalonia 2014-2020; EU Horizon 2020 FET-OPEN OPTOLogic (Grant No 899794); National Science Centre, Poland (Symfonia Grant No. 2016/20/W/ST4/00314); Marie Sk\l odowska-Curie grant STREDCH No 101029393;
“La Caixa” Junior Leaders fellowships (ID100010434) and EU Horizon 2020 under Marie Sk\l odowska-Curie grant agreement No 847648 (LCF/BQ/PI19/11690013, LCF/BQ/PI20/11760031, LCF/BQ/PR20/11770012, LCF/BQ/PR21/11840013).
P.T.G. acknowledges the Polish National Science Center Grants 2018/31/N/ST2/01429, 2020/36/T/ST2/00065 and is supported by the Foundation for Polish Science (FNP).
Center for Theoretical Physics of the Polish Academy of Sciences is a member of the National Laboratory of Atomic, Molecular and Optical Physics (KL FAMO).
A.P. and J.A. acknowledge support from the Spanish Ministry of Economy and Competitiveness through the “Severo Ochoa” Programme for Centres of Excellence in R$\&$D (Grant No. SEV-2015-0496), SuMaTe project (RTI2018-095853-B-C21) cofinanced by the European Regional Development Fund and from the Catalan Government 2017-SGR-1519. EU COST action NANOCOHYBRI CA16218. M. C. acknowledges the Guangdong Province Science and Technology Major Project (Future functional materials under extreme conditions - 212019071820400001). We also acknowledge the Scientific Services at ICMAB and ICN2.

\subsection{Authors' contributions:} 
J.B. conceived the project and supervised the experimental work. U.E., performed the measurements with support from J.B., T.P.H.S., and T.S.. U.E. and J.B. analysed the data. J.B. performed electronic structure calculations. A.P. and J.A. provided the YBCO samples and their characterisation. U.E., J.B., I.T., M.C. and M.L. contributed to the interpretation of data. U.B., M.C., T.G., P.T.G. and M.L. developed the theory and carried out the simulations. J.B. wrote the manuscript with help from U.B., U.E., and T.G.. All authors read the manuscript.

	\nocite{apsrev41Control}
\bibliography{HighTc,revtex-custom}

\providecommand{\noopsort}[1]{}\providecommand{\singleletter}[1]{#1}%
\begin{thebibliography}{42}%
\makeatletter
\providecommand \@ifxundefined [1]{%
 \@ifx{#1\undefined}
}%
\providecommand \@ifnum [1]{%
 \ifnum #1\expandafter \@firstoftwo
 \else \expandafter \@secondoftwo
 \fi
}%
\providecommand \@ifx [1]{%
 \ifx #1\expandafter \@firstoftwo
 \else \expandafter \@secondoftwo
 \fi
}%
\providecommand \natexlab [1]{#1}%
\providecommand \enquote  [1]{``#1''}%
\providecommand \bibnamefont  [1]{#1}%
\providecommand \bibfnamefont [1]{#1}%
\providecommand \citenamefont [1]{#1}%
\providecommand \href@noop [0]{\@secondoftwo}%
\providecommand \href [0]{\begingroup \@sanitize@url \@href}%
\providecommand \@href[1]{\@@startlink{#1}\@@href}%
\providecommand \@@href[1]{\endgroup#1\@@endlink}%
\providecommand \@sanitize@url [0]{\catcode `\\12\catcode `\$12\catcode
  `\&12\catcode `\#12\catcode `\^12\catcode `\_12\catcode `\%12\relax}%
\providecommand \@@startlink[1]{}%
\providecommand \@@endlink[0]{}%
\providecommand \url  [0]{\begingroup\@sanitize@url \@url }%
\providecommand \@url [1]{\endgroup\@href {#1}{\urlprefix }}%
\providecommand \urlprefix  [0]{URL }%
\providecommand \Eprint [0]{\href }%
\providecommand \doibase [0]{http://dx.doi.org/}%
\providecommand \selectlanguage [0]{\@gobble}%
\providecommand \bibinfo  [0]{\@secondoftwo}%
\providecommand \bibfield  [0]{\@secondoftwo}%
\providecommand \translation [1]{[#1]}%
\providecommand \BibitemOpen [0]{}%
\providecommand \bibitemStop [0]{}%
\providecommand \bibitemNoStop [0]{.\EOS\space}%
\providecommand \EOS [0]{\spacefactor3000\relax}%
\providecommand \BibitemShut  [1]{\csname bibitem#1\endcsname}%
\let\auto@bib@innerbib\@empty
\bibitem [{\citenamefont {Krausz}\ and\ \citenamefont
  {Ivanov}(2009)}]{Krausz2009}%
  \BibitemOpen
  \bibfield  {author} {\bibinfo {author} {\bibfnamefont {F.}~\bibnamefont
  {Krausz}}\ and\ \bibinfo {author} {\bibfnamefont {M.~Y.}\ \bibnamefont
  {Ivanov}},\ }\bibfield  {title} {\enquote {\bibinfo {title} {Attosecond
  physics},}\ }\href {\doibase 10.1103/RevModPhys.81.163} {\bibfield  {journal}
  {\bibinfo  {journal} {Rev. Mod. Phys}\ }\textbf {\bibinfo {volume} {81}},\
  \bibinfo {pages} {163--234} (\bibinfo {year} {2009})}\BibitemShut {NoStop}%
\bibitem [{\citenamefont {McPherson}\ \emph {et~al.}(1987)\citenamefont
  {McPherson}, \citenamefont {Gibson}, \citenamefont {Jara}, \citenamefont
  {Johann}, \citenamefont {Luk}, \citenamefont {McIntyre}, \citenamefont
  {Boyer},\ and\ \citenamefont {Rhodes}}]{McPherson1987}%
  \BibitemOpen
  \bibfield  {author} {\bibinfo {author} {\bibfnamefont {A.}~\bibnamefont
  {McPherson}}, \bibinfo {author} {\bibfnamefont {G.}~\bibnamefont {Gibson}},
  \bibinfo {author} {\bibfnamefont {H.}~\bibnamefont {Jara}}, \bibinfo {author}
  {\bibfnamefont {U.}~\bibnamefont {Johann}}, \bibinfo {author} {\bibfnamefont
  {T.~S.}\ \bibnamefont {Luk}}, \bibinfo {author} {\bibfnamefont {I.~A.}\
  \bibnamefont {McIntyre}}, \bibinfo {author} {\bibfnamefont {K.}~\bibnamefont
  {Boyer}}, \ and\ \bibinfo {author} {\bibfnamefont {C.~K.}\ \bibnamefont
  {Rhodes}},\ }\bibfield  {title} {\enquote {\bibinfo {title} {{Studies of
  multiphoton production of vacuum-ultraviolet radiation in the rare gases}},}\
  }\href {\doibase 10.1364/JOSAB.4.000595} {\bibfield  {journal} {\bibinfo
  {journal} {Journal of the Optical Society of America B}\ }\textbf {\bibinfo
  {volume} {4}},\ \bibinfo {pages} {595} (\bibinfo {year} {1987})}\BibitemShut
  {NoStop}%
\bibitem [{\citenamefont {Ferray}\ \emph {et~al.}(1988)\citenamefont {Ferray},
  \citenamefont {L'Huillier}, \citenamefont {Li}, \citenamefont {Lompre},
  \citenamefont {Mainfray},\ and\ \citenamefont {Manus}}]{Ferray1988}%
  \BibitemOpen
  \bibfield  {author} {\bibinfo {author} {\bibfnamefont {M.}~\bibnamefont
  {Ferray}}, \bibinfo {author} {\bibfnamefont {A.}~\bibnamefont {L'Huillier}},
  \bibinfo {author} {\bibfnamefont {X.~F.}\ \bibnamefont {Li}}, \bibinfo
  {author} {\bibfnamefont {L.~A.}\ \bibnamefont {Lompre}}, \bibinfo {author}
  {\bibfnamefont {G.}~\bibnamefont {Mainfray}}, \ and\ \bibinfo {author}
  {\bibfnamefont {C.}~\bibnamefont {Manus}},\ }\bibfield  {title} {\enquote
  {\bibinfo {title} {Multiple-harmonic conversion of 1064 nm radiation in rare
  gases},}\ }\href {\doibase 10.1088/0953-4075/21/3/001} {\bibfield  {journal}
  {\bibinfo  {journal} {J. Phys. B: At. Mol. Opt. Phys.}\ }\textbf {\bibinfo
  {volume} {21}},\ \bibinfo {pages} {L31--L35} (\bibinfo {year}
  {1988})}\BibitemShut {NoStop}%
\bibitem [{\citenamefont {Corkum}(1993)}]{Corkum1993}%
  \BibitemOpen
  \bibfield  {author} {\bibinfo {author} {\bibfnamefont {P.~B.}\ \bibnamefont
  {Corkum}},\ }\bibfield  {title} {\enquote {\bibinfo {title} {Plasma
  perspective on strong field multiphoton ionization},}\ }\href {\doibase
  10.1103/PhysRevLett.71.1994} {\bibfield  {journal} {\bibinfo  {journal}
  {Phys. Rev. Lett.}\ }\textbf {\bibinfo {volume} {71}},\ \bibinfo {pages}
  {1994--1997} (\bibinfo {year} {1993})}\BibitemShut {NoStop}%
\bibitem [{\citenamefont {Kulander}\ \emph {et~al.}(1993)\citenamefont
  {Kulander}, \citenamefont {Schafer},\ and\ \citenamefont
  {Krause}}]{Kulander1993}%
  \BibitemOpen
  \bibfield  {author} {\bibinfo {author} {\bibfnamefont {K.~C.}\ \bibnamefont
  {Kulander}}, \bibinfo {author} {\bibfnamefont {K.~J.}\ \bibnamefont
  {Schafer}}, \ and\ \bibinfo {author} {\bibfnamefont {J.~L.}\ \bibnamefont
  {Krause}},\ }\enquote {\bibinfo {title} {Dynamics of short-pulse excitation,
  ionization and harmonic conversion},}\ in\ \href {\doibase
  10.1007/978-1-4615-7963-2_10} {\emph {\bibinfo {booktitle} {Super-Intense
  Laser-Atom Physics}}},\ \bibinfo {editor} {edited by\ \bibinfo {editor}
  {\bibfnamefont {B.}~\bibnamefont {Piraux}}, \bibinfo {editor} {\bibfnamefont
  {A.}~\bibnamefont {L'Huillier}}, \ and\ \bibinfo {editor} {\bibfnamefont
  {K.}~\bibnamefont {Rz{\k{a}}{\.{z}}ewski}}}\ (\bibinfo  {publisher} {Springer
  US},\ \bibinfo {address} {Boston, MA},\ \bibinfo {year} {1993})\ pp.\
  \bibinfo {pages} {95--110}\BibitemShut {NoStop}%
\bibitem [{\citenamefont {Lewenstein}\ \emph {et~al.}(1994)\citenamefont
  {Lewenstein}, \citenamefont {Balcou}, \citenamefont {Ivanov}, \citenamefont
  {L'Huillier},\ and\ \citenamefont {Corkum}}]{Maciej1994}%
  \BibitemOpen
  \bibfield  {author} {\bibinfo {author} {\bibfnamefont {M.}~\bibnamefont
  {Lewenstein}}, \bibinfo {author} {\bibfnamefont {P.}~\bibnamefont {Balcou}},
  \bibinfo {author} {\bibfnamefont {M.~Y.}\ \bibnamefont {Ivanov}}, \bibinfo
  {author} {\bibfnamefont {A.}~\bibnamefont {L'Huillier}}, \ and\ \bibinfo
  {author} {\bibfnamefont {P.~B.}\ \bibnamefont {Corkum}},\ }\bibfield  {title}
  {\enquote {\bibinfo {title} {Theory of high-harmonic generation by
  low-frequency laser fields},}\ }\href {\doibase 10.1103/PhysRevA.49.2117}
  {\bibfield  {journal} {\bibinfo  {journal} {Phys. Rev. A}\ }\textbf {\bibinfo
  {volume} {49}},\ \bibinfo {pages} {2117--2132} (\bibinfo {year}
  {1994})}\BibitemShut {NoStop}%
\bibitem [{\citenamefont {Ghimire}\ \emph {et~al.}(2011)\citenamefont
  {Ghimire}, \citenamefont {Dichiara}, \citenamefont {Sistrunk}, \citenamefont
  {Agostini}, \citenamefont {Dimauro},\ and\ \citenamefont
  {Reis}}]{Ghimire2011}%
  \BibitemOpen
  \bibfield  {author} {\bibinfo {author} {\bibfnamefont {S.}~\bibnamefont
  {Ghimire}}, \bibinfo {author} {\bibfnamefont {A.~D.}\ \bibnamefont
  {Dichiara}}, \bibinfo {author} {\bibfnamefont {E.}~\bibnamefont {Sistrunk}},
  \bibinfo {author} {\bibfnamefont {P.}~\bibnamefont {Agostini}}, \bibinfo
  {author} {\bibfnamefont {L.~F.}\ \bibnamefont {Dimauro}}, \ and\ \bibinfo
  {author} {\bibfnamefont {D.~A.}\ \bibnamefont {Reis}},\ }\bibfield  {title}
  {\enquote {\bibinfo {title} {{Observation of high-order harmonic generation
  in a bulk crystal}},}\ }\href {\doibase 10.1038/nphys1847} {\bibfield
  {journal} {\bibinfo  {journal} {Nature Physics}\ }\textbf {\bibinfo {volume}
  {7}},\ \bibinfo {pages} {138--141} (\bibinfo {year} {2011})}\BibitemShut
  {NoStop}%
\bibitem [{\citenamefont {Hohenleutner}\ \emph {et~al.}(2015)\citenamefont
  {Hohenleutner}, \citenamefont {Langer}, \citenamefont {Schubert},
  \citenamefont {Knorr}, \citenamefont {Huttner}, \citenamefont {Koch},
  \citenamefont {Kira},\ and\ \citenamefont {Huber}}]{Hohenleutner2015}%
  \BibitemOpen
  \bibfield  {author} {\bibinfo {author} {\bibfnamefont {M.}~\bibnamefont
  {Hohenleutner}}, \bibinfo {author} {\bibfnamefont {F.}~\bibnamefont
  {Langer}}, \bibinfo {author} {\bibfnamefont {O.}~\bibnamefont {Schubert}},
  \bibinfo {author} {\bibfnamefont {M.}~\bibnamefont {Knorr}}, \bibinfo
  {author} {\bibfnamefont {U.}~\bibnamefont {Huttner}}, \bibinfo {author}
  {\bibfnamefont {S.~W.}\ \bibnamefont {Koch}}, \bibinfo {author}
  {\bibfnamefont {M.}~\bibnamefont {Kira}}, \ and\ \bibinfo {author}
  {\bibfnamefont {R.}~\bibnamefont {Huber}},\ }\bibfield  {title} {\enquote
  {\bibinfo {title} {{Real-time observation of interfering crystal electrons in
  high-harmonic generation.}}}\ }\href {\doibase 10.1038/nature14652}
  {\bibfield  {journal} {\bibinfo  {journal} {Nature}\ }\textbf {\bibinfo
  {volume} {523}},\ \bibinfo {pages} {572--5} (\bibinfo {year}
  {2015})}\BibitemShut {NoStop}%
\bibitem [{\citenamefont {Yoshikawa}\ \emph {et~al.}(2017)\citenamefont
  {Yoshikawa}, \citenamefont {Tamaya},\ and\ \citenamefont
  {Tanaka}}]{Yoshikawa2017}%
  \BibitemOpen
  \bibfield  {author} {\bibinfo {author} {\bibfnamefont {N.}~\bibnamefont
  {Yoshikawa}}, \bibinfo {author} {\bibfnamefont {T.}~\bibnamefont {Tamaya}}, \
  and\ \bibinfo {author} {\bibfnamefont {K.}~\bibnamefont {Tanaka}},\
  }\bibfield  {title} {\enquote {\bibinfo {title} {{Optics: High-harmonic
  generation in graphene enhanced by elliptically polarized light
  excitation}},}\ }\href {\doibase
  10.1126/SCIENCE.AAM8861/SUPPL_FILE/AAM8861_YOSHIKAWA_SM.PDF} {\bibfield
  {journal} {\bibinfo  {journal} {Science}\ }\textbf {\bibinfo {volume}
  {356}},\ \bibinfo {pages} {736--738} (\bibinfo {year} {2017})}\BibitemShut
  {NoStop}%
\bibitem [{\citenamefont {Baudisch}\ \emph {et~al.}(2018)\citenamefont
  {Baudisch}, \citenamefont {Marini}, \citenamefont {Cox}, \citenamefont {Zhu},
  \citenamefont {Silva}, \citenamefont {Teichmann}, \citenamefont {Massicotte},
  \citenamefont {Koppens}, \citenamefont {Levitov}, \citenamefont {Garc\'ia~de
  Abajo},\ and\ \citenamefont {Biegert}}]{Baudisch2018}%
  \BibitemOpen
  \bibfield  {author} {\bibinfo {author} {\bibfnamefont {M.}~\bibnamefont
  {Baudisch}}, \bibinfo {author} {\bibfnamefont {A.}~\bibnamefont {Marini}},
  \bibinfo {author} {\bibfnamefont {J.~D.}\ \bibnamefont {Cox}}, \bibinfo
  {author} {\bibfnamefont {T.}~\bibnamefont {Zhu}}, \bibinfo {author}
  {\bibfnamefont {F.}~\bibnamefont {Silva}}, \bibinfo {author} {\bibfnamefont
  {S.}~\bibnamefont {Teichmann}}, \bibinfo {author} {\bibfnamefont
  {M.}~\bibnamefont {Massicotte}}, \bibinfo {author} {\bibfnamefont
  {F.}~\bibnamefont {Koppens}}, \bibinfo {author} {\bibfnamefont {L.~S.}\
  \bibnamefont {Levitov}}, \bibinfo {author} {\bibfnamefont {F.~J.}\
  \bibnamefont {Garc\'ia~de Abajo}}, \ and\ \bibinfo {author} {\bibfnamefont
  {J.}~\bibnamefont {Biegert}},\ }\bibfield  {title} {\enquote {\bibinfo
  {title} {Ultrafast nonlinear optical response of dirac fermions in
  graphene},}\ }\href {\doibase 10.1038/s41467-018-03413-7} {\bibfield
  {journal} {\bibinfo  {journal} {Nat. Commun.}\ }\textbf {\bibinfo {volume}
  {9}},\ \bibinfo {pages} {1018} (\bibinfo {year} {2018})}\BibitemShut
  {NoStop}%
\bibitem [{\citenamefont {Vampa}\ \emph {et~al.}(2017)\citenamefont {Vampa},
  \citenamefont {Ghamsari}, \citenamefont {Siadat~Mousavi}, \citenamefont
  {Hammond}, \citenamefont {Olivieri}, \citenamefont {Lisicka-Skrek},
  \citenamefont {Naumov}, \citenamefont {Villeneuve}, \citenamefont {Staudte},
  \citenamefont {Berini},\ and\ \citenamefont {Corkum}}]{Vampa2017}%
  \BibitemOpen
  \bibfield  {author} {\bibinfo {author} {\bibfnamefont {G.}~\bibnamefont
  {Vampa}}, \bibinfo {author} {\bibfnamefont {B.~G.}\ \bibnamefont {Ghamsari}},
  \bibinfo {author} {\bibfnamefont {S.}~\bibnamefont {Siadat~Mousavi}},
  \bibinfo {author} {\bibfnamefont {T.~J.}\ \bibnamefont {Hammond}}, \bibinfo
  {author} {\bibfnamefont {A.}~\bibnamefont {Olivieri}}, \bibinfo {author}
  {\bibfnamefont {E.}~\bibnamefont {Lisicka-Skrek}}, \bibinfo {author}
  {\bibfnamefont {A.~Y.}\ \bibnamefont {Naumov}}, \bibinfo {author}
  {\bibfnamefont {D.~M.}\ \bibnamefont {Villeneuve}}, \bibinfo {author}
  {\bibfnamefont {A.}~\bibnamefont {Staudte}}, \bibinfo {author} {\bibfnamefont
  {P.}~\bibnamefont {Berini}}, \ and\ \bibinfo {author} {\bibfnamefont {P.~B.}\
  \bibnamefont {Corkum}},\ }\bibfield  {title} {\enquote {\bibinfo {title}
  {Plasmon-enhanced high-harmonic generation from silicon},}\ }\href {\doibase
  10.1038/nphys4087} {\bibfield  {journal} {\bibinfo  {journal} {Nat. Phys.}\
  }\textbf {\bibinfo {volume} {13}},\ \bibinfo {pages} {659--662} (\bibinfo
  {year} {2017})}\BibitemShut {NoStop}%
\bibitem [{\citenamefont {Franz}\ \emph {et~al.}(2019)\citenamefont {Franz},
  \citenamefont {Kaassamani}, \citenamefont {Gauthier}, \citenamefont
  {Nicolas}, \citenamefont {Kholodtsova}, \citenamefont {Douillard},
  \citenamefont {Gomes}, \citenamefont {Lavoute}, \citenamefont {Gaponov},
  \citenamefont {Ducros}, \citenamefont {F\'evrier}, \citenamefont {Biegert},
  \citenamefont {Shi}, \citenamefont {Kovacev}, \citenamefont {Boutu},\ and\
  \citenamefont {Merdji}}]{Franz2019}%
  \BibitemOpen
  \bibfield  {author} {\bibinfo {author} {\bibfnamefont {D.}~\bibnamefont
  {Franz}}, \bibinfo {author} {\bibfnamefont {S.}~\bibnamefont {Kaassamani}},
  \bibinfo {author} {\bibfnamefont {D.}~\bibnamefont {Gauthier}}, \bibinfo
  {author} {\bibfnamefont {R.}~\bibnamefont {Nicolas}}, \bibinfo {author}
  {\bibfnamefont {M.}~\bibnamefont {Kholodtsova}}, \bibinfo {author}
  {\bibfnamefont {L.}~\bibnamefont {Douillard}}, \bibinfo {author}
  {\bibfnamefont {J.-T.}\ \bibnamefont {Gomes}}, \bibinfo {author}
  {\bibfnamefont {L.}~\bibnamefont {Lavoute}}, \bibinfo {author} {\bibfnamefont
  {D.}~\bibnamefont {Gaponov}}, \bibinfo {author} {\bibfnamefont
  {N.}~\bibnamefont {Ducros}}, \bibinfo {author} {\bibfnamefont
  {S.}~\bibnamefont {F\'evrier}}, \bibinfo {author} {\bibfnamefont
  {J.}~\bibnamefont {Biegert}}, \bibinfo {author} {\bibfnamefont
  {L.}~\bibnamefont {Shi}}, \bibinfo {author} {\bibfnamefont {M.}~\bibnamefont
  {Kovacev}}, \bibinfo {author} {\bibfnamefont {W.}~\bibnamefont {Boutu}}, \
  and\ \bibinfo {author} {\bibfnamefont {H.}~\bibnamefont {Merdji}},\
  }\bibfield  {title} {\enquote {\bibinfo {title} {All semiconductor enhanced
  high-harmonic generation from a single nanostructured cone},}\ }\href
  {\doibase 10.1038/s41598-019-41642-y} {\bibfield  {journal} {\bibinfo
  {journal} {Sci. Rep.}\ }\textbf {\bibinfo {volume} {9}},\ \bibinfo {pages}
  {5663} (\bibinfo {year} {2019})}\BibitemShut {NoStop}%
\bibitem [{\citenamefont {Lanin}\ \emph {et~al.}(2017)\citenamefont {Lanin},
  \citenamefont {Stepanov}, \citenamefont {Fedotov}, ,\ and\ \citenamefont
  {Zheltikov}}]{Lanin2017}%
  \BibitemOpen
  \bibfield  {author} {\bibinfo {author} {\bibfnamefont {A.~A.}\ \bibnamefont
  {Lanin}}, \bibinfo {author} {\bibfnamefont {E.~A.}\ \bibnamefont {Stepanov}},
  \bibinfo {author} {\bibfnamefont {A.~B.}\ \bibnamefont {Fedotov}}, , \ and\
  \bibinfo {author} {\bibfnamefont {A.~M.}\ \bibnamefont {Zheltikov}},\
  }\bibfield  {title} {\enquote {\bibinfo {title} {Mapping the electron band
  structure by intraband high-harmonic generation in solids},}\ }\href
  {\doibase 10.1364/OPTICA.4.000516} {\bibfield  {journal} {\bibinfo  {journal}
  {Optica}\ }\textbf {\bibinfo {volume} {4}},\ \bibinfo {pages} {516--519}
  (\bibinfo {year} {2017})}\BibitemShut {NoStop}%
\bibitem [{\citenamefont {Luu}\ and\ \citenamefont
  {W{\"o}rner}(2018)}]{Luu2018}%
  \BibitemOpen
  \bibfield  {author} {\bibinfo {author} {\bibfnamefont {T.~T.}\ \bibnamefont
  {Luu}}\ and\ \bibinfo {author} {\bibfnamefont {H.~J.}\ \bibnamefont
  {W{\"o}rner}},\ }\bibfield  {title} {\enquote {\bibinfo {title} {Measurement
  of the berry curvature of solids using high-harmonic spectroscopy},}\ }\href
  {\doibase 10.1038/s41467-018-03397-4} {\bibfield  {journal} {\bibinfo
  {journal} {Nat. Commun.}\ }\textbf {\bibinfo {volume} {9}},\ \bibinfo {pages}
  {916} (\bibinfo {year} {2018})}\BibitemShut {NoStop}%
\bibitem [{\citenamefont {Silva}\ \emph {et~al.}(2019)\citenamefont {Silva},
  \citenamefont {Jim{\'{e}}nez-Gal{\'{a}}n}, \citenamefont {Amorim},
  \citenamefont {Smirnova},\ and\ \citenamefont {Ivanov}}]{Silva2019}%
  \BibitemOpen
  \bibfield  {author} {\bibinfo {author} {\bibfnamefont {R.~E.}\ \bibnamefont
  {Silva}}, \bibinfo {author} {\bibnamefont {Jim{\'{e}}nez-Gal{\'{a}}n}},
  \bibinfo {author} {\bibfnamefont {B.}~\bibnamefont {Amorim}}, \bibinfo
  {author} {\bibfnamefont {O.}~\bibnamefont {Smirnova}}, \ and\ \bibinfo
  {author} {\bibfnamefont {M.}~\bibnamefont {Ivanov}},\ }\bibfield  {title}
  {\enquote {\bibinfo {title} {{Topological strong-field physics on
  sub-laser-cycle timescale}},}\ }\href {\doibase 10.1038/s41566-019-0516-1}
  {\bibfield  {journal} {\bibinfo  {journal} {Nature Photonics 2019 13:12}\
  }\textbf {\bibinfo {volume} {13}},\ \bibinfo {pages} {849--854} (\bibinfo
  {year} {2019})},\ \Eprint {http://arxiv.org/abs/1806.11232} {1806.11232}
  \BibitemShut {NoStop}%
\bibitem [{\citenamefont {Maczewsky}\ \emph {et~al.}(2020)\citenamefont
  {Maczewsky}, \citenamefont {Heinrich}, \citenamefont {Kremer}, \citenamefont
  {Ivanov}, \citenamefont {Ehrhardt}, \citenamefont {Martinez}, \citenamefont
  {Kartashov}, \citenamefont {Konotop}, \citenamefont {Torner}, \citenamefont
  {Bauer},\ and\ \citenamefont {Szameit}}]{Maczewsky2020}%
  \BibitemOpen
  \bibfield  {author} {\bibinfo {author} {\bibfnamefont {L.~J.}\ \bibnamefont
  {Maczewsky}}, \bibinfo {author} {\bibfnamefont {M.}~\bibnamefont {Heinrich}},
  \bibinfo {author} {\bibfnamefont {M.}~\bibnamefont {Kremer}}, \bibinfo
  {author} {\bibfnamefont {S.~K.}\ \bibnamefont {Ivanov}}, \bibinfo {author}
  {\bibfnamefont {M.}~\bibnamefont {Ehrhardt}}, \bibinfo {author}
  {\bibfnamefont {F.}~\bibnamefont {Martinez}}, \bibinfo {author}
  {\bibfnamefont {Y.~V.}\ \bibnamefont {Kartashov}}, \bibinfo {author}
  {\bibfnamefont {V.~V.}\ \bibnamefont {Konotop}}, \bibinfo {author}
  {\bibfnamefont {L.}~\bibnamefont {Torner}}, \bibinfo {author} {\bibfnamefont
  {D.}~\bibnamefont {Bauer}}, \ and\ \bibinfo {author} {\bibfnamefont
  {A.}~\bibnamefont {Szameit}},\ }\bibfield  {title} {\enquote {\bibinfo
  {title} {{Nonlinearity-induced photonic topological insulator}},}\ }\href
  {https://www.science.org/doi/abs/10.1126/science.abd2033} {\bibfield
  {journal} {\bibinfo  {journal} {Science}\ }\textbf {\bibinfo {volume} {370}}
  (\bibinfo {year} {2020})},\ \Eprint {http://arxiv.org/abs/2010.12250}
  {arXiv:2010.12250} \BibitemShut {NoStop}%
\bibitem [{\citenamefont {Schmid}\ \emph {et~al.}(2021)\citenamefont {Schmid},
  \citenamefont {Weigl}, \citenamefont {Gr{\"{o}}ssing}, \citenamefont {Junk},
  \citenamefont {Gorini}, \citenamefont {Schlauderer}, \citenamefont {Ito},
  \citenamefont {Meierhofer}, \citenamefont {Hofmann}, \citenamefont
  {Afanasiev}, \citenamefont {Crewse}, \citenamefont {Kokh}, \citenamefont
  {Tereshchenko}, \citenamefont {G{\"{u}}dde}, \citenamefont {Evers},
  \citenamefont {Wilhelm}, \citenamefont {Richter}, \citenamefont
  {H{\"{o}}fer},\ and\ \citenamefont {Huber}}]{Schmid2021}%
  \BibitemOpen
  \bibfield  {author} {\bibinfo {author} {\bibfnamefont {C.~P.}\ \bibnamefont
  {Schmid}}, \bibinfo {author} {\bibfnamefont {L.}~\bibnamefont {Weigl}},
  \bibinfo {author} {\bibfnamefont {P.}~\bibnamefont {Gr{\"{o}}ssing}},
  \bibinfo {author} {\bibfnamefont {V.}~\bibnamefont {Junk}}, \bibinfo {author}
  {\bibfnamefont {C.}~\bibnamefont {Gorini}}, \bibinfo {author} {\bibfnamefont
  {S.}~\bibnamefont {Schlauderer}}, \bibinfo {author} {\bibfnamefont
  {S.}~\bibnamefont {Ito}}, \bibinfo {author} {\bibfnamefont {M.}~\bibnamefont
  {Meierhofer}}, \bibinfo {author} {\bibfnamefont {N.}~\bibnamefont {Hofmann}},
  \bibinfo {author} {\bibfnamefont {D.}~\bibnamefont {Afanasiev}}, \bibinfo
  {author} {\bibfnamefont {J.}~\bibnamefont {Crewse}}, \bibinfo {author}
  {\bibfnamefont {K.~A.}\ \bibnamefont {Kokh}}, \bibinfo {author}
  {\bibfnamefont {O.~E.}\ \bibnamefont {Tereshchenko}}, \bibinfo {author}
  {\bibfnamefont {J.}~\bibnamefont {G{\"{u}}dde}}, \bibinfo {author}
  {\bibfnamefont {F.}~\bibnamefont {Evers}}, \bibinfo {author} {\bibfnamefont
  {J.}~\bibnamefont {Wilhelm}}, \bibinfo {author} {\bibfnamefont
  {K.}~\bibnamefont {Richter}}, \bibinfo {author} {\bibfnamefont
  {U.}~\bibnamefont {H{\"{o}}fer}}, \ and\ \bibinfo {author} {\bibfnamefont
  {R.}~\bibnamefont {Huber}},\ }\bibfield  {title} {\enquote {\bibinfo {title}
  {{Tunable non-integer high-harmonic generation in a topological
  insulator}},}\ }\href {\doibase 10.1038/s41586-021-03466-7} {\bibfield
  {journal} {\bibinfo  {journal} {Nature 2021 593:7859}\ }\textbf {\bibinfo
  {volume} {593}},\ \bibinfo {pages} {385--390} (\bibinfo {year}
  {2021})}\BibitemShut {NoStop}%
\bibitem [{\citenamefont {Baykusheva}\ \emph {et~al.}(2021)\citenamefont
  {Baykusheva}, \citenamefont {Chac{\'{o}}n}, \citenamefont {Lu}, \citenamefont
  {Bailey}, \citenamefont {Sobota}, \citenamefont {Soifer}, \citenamefont
  {Kirchmann}, \citenamefont {Rotundu}, \citenamefont {Uher}, \citenamefont
  {Heinz}, \citenamefont {Reis},\ and\ \citenamefont
  {Ghimire}}]{Baykusheva2021}%
  \BibitemOpen
  \bibfield  {author} {\bibinfo {author} {\bibfnamefont {D.}~\bibnamefont
  {Baykusheva}}, \bibinfo {author} {\bibfnamefont {A.}~\bibnamefont
  {Chac{\'{o}}n}}, \bibinfo {author} {\bibfnamefont {J.}~\bibnamefont {Lu}},
  \bibinfo {author} {\bibfnamefont {T.~P.}\ \bibnamefont {Bailey}}, \bibinfo
  {author} {\bibfnamefont {J.~A.}\ \bibnamefont {Sobota}}, \bibinfo {author}
  {\bibfnamefont {H.}~\bibnamefont {Soifer}}, \bibinfo {author} {\bibfnamefont
  {P.~S.}\ \bibnamefont {Kirchmann}}, \bibinfo {author} {\bibfnamefont
  {C.}~\bibnamefont {Rotundu}}, \bibinfo {author} {\bibfnamefont
  {C.}~\bibnamefont {Uher}}, \bibinfo {author} {\bibfnamefont {T.~F.}\
  \bibnamefont {Heinz}}, \bibinfo {author} {\bibfnamefont {D.~A.}\ \bibnamefont
  {Reis}}, \ and\ \bibinfo {author} {\bibfnamefont {S.}~\bibnamefont
  {Ghimire}},\ }\bibfield  {title} {\enquote {\bibinfo {title} {{All-Optical
  Probe of Three-Dimensional Topological Insulators Based on High-Harmonic
  Generation by Circularly Polarized Laser Fields}},}\ }\href {\doibase
  10.1021/ACS.NANOLETT.1C02145/SUPPL_FILE/NL1C02145_SI_001.PDF} {\bibfield
  {journal} {\bibinfo  {journal} {Nano Letters}\ }\textbf {\bibinfo {volume}
  {21}},\ \bibinfo {pages} {8970--8978} (\bibinfo {year} {2021})},\ \Eprint
  {http://arxiv.org/abs/2109.15291} {arXiv:2109.15291} \BibitemShut {NoStop}%
\bibitem [{\citenamefont {Stephen}(1965)}]{Stephen1965}%
  \BibitemOpen
  \bibfield  {author} {\bibinfo {author} {\bibfnamefont {M.~J.}\ \bibnamefont
  {Stephen}},\ }\bibfield  {title} {\enquote {\bibinfo {title} {Transport
  equations for superconductors},}\ }\href {\doibase 10.1103/PhysRev.139.A197}
  {\bibfield  {journal} {\bibinfo  {journal} {Phys. Rev.}\ }\textbf {\bibinfo
  {volume} {139}} (\bibinfo {year} {1965}),\
  10.1103/PhysRev.139.A197}\BibitemShut {NoStop}%
\bibitem [{\citenamefont {Ginzburg}\ and\ \citenamefont
  {Landau}(1965)}]{Ginzburg1965}%
  \BibitemOpen
  \bibfield  {author} {\bibinfo {author} {\bibfnamefont {V.~L.}\ \bibnamefont
  {Ginzburg}}\ and\ \bibinfo {author} {\bibfnamefont {L.~D.}\ \bibnamefont
  {Landau}},\ }\bibfield  {title} {\enquote {\bibinfo {title} {To the theory of
  superconductivity},}\ }\href@noop {} {\bibfield  {journal} {\bibinfo
  {journal} {Zh. Eksp. Teor. Fiz.}\ }\textbf {\bibinfo {volume} {20}},\
  \bibinfo {pages} {1064} (\bibinfo {year} {1965})}\BibitemShut {NoStop}%
\bibitem [{\citenamefont {Barankov}\ and\ \citenamefont
  {Levitov}(2006)}]{Barankov2006}%
  \BibitemOpen
  \bibfield  {author} {\bibinfo {author} {\bibfnamefont {R.~A.}\ \bibnamefont
  {Barankov}}\ and\ \bibinfo {author} {\bibfnamefont {L.~S.}\ \bibnamefont
  {Levitov}},\ }\bibfield  {title} {\enquote {\bibinfo {title} {Synchronization
  in the bcs pairing dynamics as a critical phenomenon},}\ }\href {\doibase
  10.1103/PhysRevLett.96.230403} {\bibfield  {journal} {\bibinfo  {journal}
  {Phys. Rev. Lett.}\ }\textbf {\bibinfo {volume} {96}} (\bibinfo {year}
  {2006}),\ 10.1103/PhysRevLett.96.230403}\BibitemShut {NoStop}%
\bibitem [{\citenamefont {Yuzbashyan}\ \emph {et~al.}(2006)\citenamefont
  {Yuzbashyan}, \citenamefont {Tsyplyatyev},\ and\ \citenamefont
  {Altshuler}}]{Yuzbashyan2006}%
  \BibitemOpen
  \bibfield  {author} {\bibinfo {author} {\bibfnamefont {E.~A.}\ \bibnamefont
  {Yuzbashyan}}, \bibinfo {author} {\bibfnamefont {O.}~\bibnamefont
  {Tsyplyatyev}}, \ and\ \bibinfo {author} {\bibfnamefont {B.~L.}\ \bibnamefont
  {Altshuler}},\ }\bibfield  {title} {\enquote {\bibinfo {title} {Relaxation
  and persistent oscillations of the order parameter in fermionic
  condensates},}\ }\href {\doibase 10.1103/PhysRevLett.96.097005} {\bibfield
  {journal} {\bibinfo  {journal} {Phys. Rev. Lett.}\ }\textbf {\bibinfo
  {volume} {96}} (\bibinfo {year} {2006}),\
  10.1103/PhysRevLett.96.097005}\BibitemShut {NoStop}%
\bibitem [{\citenamefont {Seibold}\ \emph {et~al.}(2021)\citenamefont
  {Seibold}, \citenamefont {Udina}, \citenamefont {Castellani},\ and\
  \citenamefont {Benfatto}}]{Seibold2021}%
  \BibitemOpen
  \bibfield  {author} {\bibinfo {author} {\bibfnamefont {G.}~\bibnamefont
  {Seibold}}, \bibinfo {author} {\bibfnamefont {M.}~\bibnamefont {Udina}},
  \bibinfo {author} {\bibfnamefont {C.}~\bibnamefont {Castellani}}, \ and\
  \bibinfo {author} {\bibfnamefont {L.}~\bibnamefont {Benfatto}},\ }\bibfield
  {title} {\enquote {\bibinfo {title} {{Third harmonic generation from
  collective modes in disordered superconductors}},}\ }\href {\doibase
  10.1103/PHYSREVB.103.014512/FIGURES/13/MEDIUM} {\bibfield  {journal}
  {\bibinfo  {journal} {Physical Review B}\ }\textbf {\bibinfo {volume}
  {103}},\ \bibinfo {pages} {014512} (\bibinfo {year} {2021})},\ \Eprint
  {http://arxiv.org/abs/2010.12507} {arXiv:2010.12507} \BibitemShut {NoStop}%
\bibitem [{\citenamefont {Cot{\'{o}}n}\ \emph {et~al.}(2010)\citenamefont
  {Cot{\'{o}}n}, \citenamefont {Guzm{\'{a}}n}, \citenamefont {Ramallo},
  \citenamefont {R{\'{i}}os}, \citenamefont {Torr{\'{o}}n},\ and\ \citenamefont
  {Vidal}}]{Coton2010}%
  \BibitemOpen
  \bibfield  {author} {\bibinfo {author} {\bibfnamefont {N.}~\bibnamefont
  {Cot{\'{o}}n}}, \bibinfo {author} {\bibfnamefont {F.~J.}\ \bibnamefont
  {Guzm{\'{a}}n}}, \bibinfo {author} {\bibfnamefont {M.~V.}\ \bibnamefont
  {Ramallo}}, \bibinfo {author} {\bibfnamefont {A.}~\bibnamefont {R{\'{i}}os}},
  \bibinfo {author} {\bibfnamefont {C.}~\bibnamefont {Torr{\'{o}}n}}, \ and\
  \bibinfo {author} {\bibfnamefont {F.}~\bibnamefont {Vidal}},\ }\bibfield
  {title} {\enquote {\bibinfo {title} {{Thermal fluctuations near a phase
  transition probed through the electrical resistivity of high-temperature
  superconductors}},}\ }\href {\doibase 10.1119/1.3276713} {\bibfield
  {journal} {\bibinfo  {journal} {American Journal of Physics}\ }\textbf
  {\bibinfo {volume} {78}},\ \bibinfo {pages} {310} (\bibinfo {year}
  {2010})}\BibitemShut {NoStop}%
\bibitem [{\citenamefont {Solov'Ev}\ and\ \citenamefont
  {Dmitriev}(2009)}]{SolovEv2009}%
  \BibitemOpen
  \bibfield  {author} {\bibinfo {author} {\bibfnamefont {A.~L.}\ \bibnamefont
  {Solov'Ev}}\ and\ \bibinfo {author} {\bibfnamefont {V.~M.}\ \bibnamefont
  {Dmitriev}},\ }\bibfield  {title} {\enquote {\bibinfo {title} {{Fluctuation
  conductivity and pseudogap in YBCO high-temperature superconductors
  (Review)}},}\ }\href {\doibase 10.1063/1.3081150} {\bibfield  {journal}
  {\bibinfo  {journal} {Low Temperature Physics}\ }\textbf {\bibinfo {volume}
  {35}},\ \bibinfo {pages} {169} (\bibinfo {year} {2009})}\BibitemShut
  {NoStop}%
\bibitem [{\citenamefont {Sigrist}\ and\ \citenamefont
  {Ueda}(1991)}]{Sigrist1991}%
  \BibitemOpen
  \bibfield  {author} {\bibinfo {author} {\bibfnamefont {M.}~\bibnamefont
  {Sigrist}}\ and\ \bibinfo {author} {\bibfnamefont {K.}~\bibnamefont {Ueda}},\
  }\bibfield  {title} {\enquote {\bibinfo {title} {{Phenomenological theory of
  unconventional superconductivity}},}\ }\href {\doibase
  10.1103/RevModPhys.63.239} {\bibfield  {journal} {\bibinfo  {journal}
  {Reviews of Modern Physics}\ }\textbf {\bibinfo {volume} {63}},\ \bibinfo
  {pages} {239} (\bibinfo {year} {1991})}\BibitemShut {NoStop}%
\bibitem [{\citenamefont {Gauquelin}\ \emph {et~al.}(2014)\citenamefont
  {Gauquelin}, \citenamefont {Hawthorn}, \citenamefont {Sawatzky},
  \citenamefont {Liang}, \citenamefont {Bonn}, \citenamefont {Hardy},\ and\
  \citenamefont {Botton}}]{Gauquelin2014}%
  \BibitemOpen
  \bibfield  {author} {\bibinfo {author} {\bibfnamefont {N.}~\bibnamefont
  {Gauquelin}}, \bibinfo {author} {\bibfnamefont {D.~G.}\ \bibnamefont
  {Hawthorn}}, \bibinfo {author} {\bibfnamefont {G.~A.}\ \bibnamefont
  {Sawatzky}}, \bibinfo {author} {\bibfnamefont {R.~X.}\ \bibnamefont {Liang}},
  \bibinfo {author} {\bibfnamefont {D.~A.}\ \bibnamefont {Bonn}}, \bibinfo
  {author} {\bibfnamefont {W.~N.}\ \bibnamefont {Hardy}}, \ and\ \bibinfo
  {author} {\bibfnamefont {G.~A.}\ \bibnamefont {Botton}},\ }\bibfield  {title}
  {\enquote {\bibinfo {title} {{Atomic scale real-space mapping of holes in
  YBa2Cu3O6+$\delta$}},}\ }\href {\doibase 10.1038/ncomms5275} {\bibfield
  {journal} {\bibinfo  {journal} {Nature Communications}\ }\textbf {\bibinfo
  {volume} {5}},\ \bibinfo {pages} {1--7} (\bibinfo {year} {2014})}\BibitemShut
  {NoStop}%
\bibitem [{\citenamefont {Saxena}(1992)}]{Saxena2012}%
  \BibitemOpen
  \bibfield  {author} {\bibinfo {author} {\bibfnamefont {A.~K.}\ \bibnamefont
  {Saxena}},\ }\href {\doibase 10.1007/978-3-642-28481-6} {\emph {\bibinfo
  {title} {High-Temperature Superconductors}}}\ (\bibinfo  {publisher}
  {Springer-Verlag},\ \bibinfo {address} {Berlin Heidelberg},\ \bibinfo {year}
  {1992})\BibitemShut {NoStop}%
\bibitem [{\citenamefont {Hirao}\ \emph {et~al.}(1992)\citenamefont {Hirao},
  \citenamefont {Uda},\ and\ \citenamefont {Murayama}}]{Hirao1992}%
  \BibitemOpen
  \bibfield  {author} {\bibinfo {author} {\bibfnamefont {M.}~\bibnamefont
  {Hirao}}, \bibinfo {author} {\bibfnamefont {T.}~\bibnamefont {Uda}}, \ and\
  \bibinfo {author} {\bibfnamefont {Y.}~\bibnamefont {Murayama}},\ }\bibfield
  {title} {\enquote {\bibinfo {title} {Electronic band structure of
  {YB}a$_2${C}u$_3${O}$_7$ and {YB}a$_2${C}u$_3${O}$_6$ by density functional
  pseudopotential method},}\ }\href {\doibase 10.1016/0921-4534(92)90345-D}
  {\bibfield  {journal} {\bibinfo  {journal} {Physica C}\ }\textbf {\bibinfo
  {volume} {195}},\ \bibinfo {pages} {230--238} (\bibinfo {year}
  {1992})}\BibitemShut {NoStop}%
\bibitem [{\citenamefont {Elu}\ \emph {et~al.}(2017)\citenamefont {Elu},
  \citenamefont {Baudisch}, \citenamefont {Pires}, \citenamefont {Tani},
  \citenamefont {Frosz}, \citenamefont {K\"{o}ttig}, \citenamefont {Ermolov},
  \citenamefont {Russell},\ and\ \citenamefont {Biegert}}]{Ugaitz2017}%
  \BibitemOpen
  \bibfield  {author} {\bibinfo {author} {\bibfnamefont {U.}~\bibnamefont
  {Elu}}, \bibinfo {author} {\bibfnamefont {M.}~\bibnamefont {Baudisch}},
  \bibinfo {author} {\bibfnamefont {H.}~\bibnamefont {Pires}}, \bibinfo
  {author} {\bibfnamefont {F.}~\bibnamefont {Tani}}, \bibinfo {author}
  {\bibfnamefont {M.~H.}\ \bibnamefont {Frosz}}, \bibinfo {author}
  {\bibfnamefont {F.}~\bibnamefont {K\"{o}ttig}}, \bibinfo {author}
  {\bibfnamefont {A.}~\bibnamefont {Ermolov}}, \bibinfo {author} {\bibfnamefont
  {P.~S.}\ \bibnamefont {Russell}}, \ and\ \bibinfo {author} {\bibfnamefont
  {J.}~\bibnamefont {Biegert}},\ }\bibfield  {title} {\enquote {\bibinfo
  {title} {High average power and single-cycle pulses from a mid-ir optical
  parametric chirped pulse amplifier},}\ }\href {\doibase
  10.1364/OPTICA.4.001024} {\bibfield  {journal} {\bibinfo  {journal} {Optica}\
  }\textbf {\bibinfo {volume} {4}},\ \bibinfo {pages} {1024--1029} (\bibinfo
  {year} {2017})}\BibitemShut {NoStop}%
\bibitem [{\citenamefont {Elu}\ \emph {et~al.}(2020)\citenamefont {Elu},
  \citenamefont {Maidment}, \citenamefont {Vamos}, \citenamefont {Tani},
  \citenamefont {Novoa}, \citenamefont {Frosz}, \citenamefont {Badikov},
  \citenamefont {Badikov}, \citenamefont {Petrov}, \citenamefont {{St. J.
  Russell}},\ and\ \citenamefont {Biegert}}]{Elu2020}%
  \BibitemOpen
  \bibfield  {author} {\bibinfo {author} {\bibfnamefont {U.}~\bibnamefont
  {Elu}}, \bibinfo {author} {\bibfnamefont {L.}~\bibnamefont {Maidment}},
  \bibinfo {author} {\bibfnamefont {L.}~\bibnamefont {Vamos}}, \bibinfo
  {author} {\bibfnamefont {F.}~\bibnamefont {Tani}}, \bibinfo {author}
  {\bibfnamefont {D.}~\bibnamefont {Novoa}}, \bibinfo {author} {\bibfnamefont
  {M.~H.}\ \bibnamefont {Frosz}}, \bibinfo {author} {\bibfnamefont
  {V.}~\bibnamefont {Badikov}}, \bibinfo {author} {\bibfnamefont
  {D.}~\bibnamefont {Badikov}}, \bibinfo {author} {\bibfnamefont
  {V.}~\bibnamefont {Petrov}}, \bibinfo {author} {\bibfnamefont
  {P.}~\bibnamefont {{St. J. Russell}}}, \ and\ \bibinfo {author}
  {\bibfnamefont {J.}~\bibnamefont {Biegert}},\ }\bibfield  {title} {\enquote
  {\bibinfo {title} {{Seven-octave high-brightness and
  carrier-envelope-phase-stable light source}},}\ }\href {\doibase
  10.1038/s41566-020-00735-1} {\bibfield  {journal} {\bibinfo  {journal}
  {Nature Photonics 2020 15:4}\ }\textbf {\bibinfo {volume} {15}},\ \bibinfo
  {pages} {277--280} (\bibinfo {year} {2020})}\BibitemShut {NoStop}%
\bibitem [{\citenamefont {Hubbard}(1963)}]{Hubbard1963}%
  \BibitemOpen
  \bibfield  {author} {\bibinfo {author} {\bibfnamefont {J.}~\bibnamefont
  {Hubbard}},\ }\bibfield  {title} {\enquote {\bibinfo {title} {Electron
  correlations in narrow energy bands},}\ }\href {\doibase
  10.1098/rspa.1963.0204} {\bibfield  {journal} {\bibinfo  {journal} {Proc.
  Roy. Soc. London A}\ }\textbf {\bibinfo {volume} {276}},\ \bibinfo {pages}
  {238} (\bibinfo {year} {1963})}\BibitemShut {NoStop}%
\bibitem [{\citenamefont {Emery}(1987)}]{Emery1987}%
  \BibitemOpen
  \bibfield  {author} {\bibinfo {author} {\bibfnamefont {V.~J.}\ \bibnamefont
  {Emery}},\ }\bibfield  {title} {\enquote {\bibinfo {title} {Theory of
  high-${\mathrm{t}}_{\mathrm{c}}$ superconductivity in oxides},}\ }\href
  {\doibase 10.1103/PhysRevLett.58.2794} {\bibfield  {journal} {\bibinfo
  {journal} {Phys. Rev. Lett.}\ }\textbf {\bibinfo {volume} {58}},\ \bibinfo
  {pages} {2794--2797} (\bibinfo {year} {1987})}\BibitemShut {NoStop}%
\bibitem [{\citenamefont {Zhang}\ and\ \citenamefont {Rice}(1988)}]{Zhang1988}%
  \BibitemOpen
  \bibfield  {author} {\bibinfo {author} {\bibfnamefont {F.~C.}\ \bibnamefont
  {Zhang}}\ and\ \bibinfo {author} {\bibfnamefont {T.~M.}\ \bibnamefont
  {Rice}},\ }\bibfield  {title} {\enquote {\bibinfo {title} {Effective
  hamiltonian for the superconducting cu oxides},}\ }\href {\doibase
  10.1103/PhysRevB.37.3759} {\bibfield  {journal} {\bibinfo  {journal} {Phys.
  Rev. B}\ }\textbf {\bibinfo {volume} {37}},\ \bibinfo {pages} {3759--3761}
  (\bibinfo {year} {1988})}\BibitemShut {NoStop}%
\bibitem [{\citenamefont {Hirsch}(1987)}]{Hirsch1987}%
  \BibitemOpen
  \bibfield  {author} {\bibinfo {author} {\bibfnamefont {J.~E.}\ \bibnamefont
  {Hirsch}},\ }\bibfield  {title} {\enquote {\bibinfo {title}
  {Antiferromagnetism, localization, and pairing in a two-dimensional model for
  ${\mathrm{cuo}}_{2}$},}\ }\href {\doibase 10.1103/PhysRevLett.59.228}
  {\bibfield  {journal} {\bibinfo  {journal} {Phys. Rev. Lett.}\ }\textbf
  {\bibinfo {volume} {59}},\ \bibinfo {pages} {228--231} (\bibinfo {year}
  {1987})}\BibitemShut {NoStop}%
\bibitem [{\citenamefont {Kung}\ \emph {et~al.}(2016)\citenamefont {Kung},
  \citenamefont {Chen}, \citenamefont {Wang}, \citenamefont {Huang},
  \citenamefont {Nowadnick}, \citenamefont {Moritz}, \citenamefont {Scalettar},
  \citenamefont {Johnston},\ and\ \citenamefont {Devereaux}}]{Kung2016}%
  \BibitemOpen
  \bibfield  {author} {\bibinfo {author} {\bibfnamefont {Y.~F.}\ \bibnamefont
  {Kung}}, \bibinfo {author} {\bibfnamefont {C.-C.}\ \bibnamefont {Chen}},
  \bibinfo {author} {\bibfnamefont {Y.}~\bibnamefont {Wang}}, \bibinfo {author}
  {\bibfnamefont {E.~W.}\ \bibnamefont {Huang}}, \bibinfo {author}
  {\bibfnamefont {E.~A.}\ \bibnamefont {Nowadnick}}, \bibinfo {author}
  {\bibfnamefont {B.}~\bibnamefont {Moritz}}, \bibinfo {author} {\bibfnamefont
  {R.~T.}\ \bibnamefont {Scalettar}}, \bibinfo {author} {\bibfnamefont
  {S.}~\bibnamefont {Johnston}}, \ and\ \bibinfo {author} {\bibfnamefont
  {T.~P.}\ \bibnamefont {Devereaux}},\ }\bibfield  {title} {\enquote {\bibinfo
  {title} {Characterizing the three-orbital hubbard model with determinant
  quantum monte carlo},}\ }\href {\doibase 10.1103/PhysRevB.93.155166}
  {\bibfield  {journal} {\bibinfo  {journal} {Phys. Rev. B}\ }\textbf {\bibinfo
  {volume} {93}},\ \bibinfo {pages} {155166} (\bibinfo {year}
  {2016})}\BibitemShut {NoStop}%
\bibitem [{\citenamefont {Scalettar}\ \emph {et~al.}(1991)\citenamefont
  {Scalettar}, \citenamefont {Scalapino}, \citenamefont {Sugar},\ and\
  \citenamefont {White}}]{Scalettar1991}%
  \BibitemOpen
  \bibfield  {author} {\bibinfo {author} {\bibfnamefont {R.~T.}\ \bibnamefont
  {Scalettar}}, \bibinfo {author} {\bibfnamefont {D.~J.}\ \bibnamefont
  {Scalapino}}, \bibinfo {author} {\bibfnamefont {R.~L.}\ \bibnamefont
  {Sugar}}, \ and\ \bibinfo {author} {\bibfnamefont {S.~R.}\ \bibnamefont
  {White}},\ }\bibfield  {title} {\enquote {\bibinfo {title}
  {Antiferromagnetic, charge-transfer, and pairing correlations in the
  three-band hubbard model},}\ }\href {\doibase 10.1103/PhysRevB.44.770}
  {\bibfield  {journal} {\bibinfo  {journal} {Phys. Rev. B}\ }\textbf {\bibinfo
  {volume} {44}},\ \bibinfo {pages} {770--781} (\bibinfo {year}
  {1991})}\BibitemShut {NoStop}%
\bibitem [{\citenamefont {Shekhter}\ \emph {et~al.}(2013)\citenamefont
  {Shekhter}, \citenamefont {Ramshaw}, \citenamefont {Liang}, \citenamefont
  {Hardy}, \citenamefont {Bonn}, \citenamefont {Balakirev}, \citenamefont
  {McDonald}, \citenamefont {Betts}, \citenamefont {Riggs},\ and\ \citenamefont
  {Migliori}}]{Shekhter2013}%
  \BibitemOpen
  \bibfield  {author} {\bibinfo {author} {\bibfnamefont {A.}~\bibnamefont
  {Shekhter}}, \bibinfo {author} {\bibfnamefont {B.~J.}\ \bibnamefont
  {Ramshaw}}, \bibinfo {author} {\bibfnamefont {R.}~\bibnamefont {Liang}},
  \bibinfo {author} {\bibfnamefont {W.~N.}\ \bibnamefont {Hardy}}, \bibinfo
  {author} {\bibfnamefont {D.~A.}\ \bibnamefont {Bonn}}, \bibinfo {author}
  {\bibfnamefont {F.~F.}\ \bibnamefont {Balakirev}}, \bibinfo {author}
  {\bibfnamefont {R.~D.}\ \bibnamefont {McDonald}}, \bibinfo {author}
  {\bibfnamefont {J.~B.}\ \bibnamefont {Betts}}, \bibinfo {author}
  {\bibfnamefont {S.~C.}\ \bibnamefont {Riggs}}, \ and\ \bibinfo {author}
  {\bibfnamefont {A.}~\bibnamefont {Migliori}},\ }\bibfield  {title} {\enquote
  {\bibinfo {title} {{Bounding the pseudogap with a line of phase transitions
  in YBa2Cu3O6+$\delta$}},}\ }\href {\doibase 10.1038/nature12165} {\bibfield
  {journal} {\bibinfo  {journal} {Nature 2013 498:7452}\ }\textbf {\bibinfo
  {volume} {498}},\ \bibinfo {pages} {75--77} (\bibinfo {year}
  {2013})}\BibitemShut {NoStop}%
\bibitem [{\citenamefont {Ventura}\ \emph {et~al.}(2017)\citenamefont
  {Ventura}, \citenamefont {Passos}, \citenamefont {Lopes~dos Santos},
  \citenamefont {Viana Parente~Lopes},\ and\ \citenamefont
  {Peres}}]{Ventura2017}%
  \BibitemOpen
  \bibfield  {author} {\bibinfo {author} {\bibfnamefont {G.~B.}\ \bibnamefont
  {Ventura}}, \bibinfo {author} {\bibfnamefont {D.~J.}\ \bibnamefont {Passos}},
  \bibinfo {author} {\bibfnamefont {J.~M.~B.}\ \bibnamefont {Lopes~dos
  Santos}}, \bibinfo {author} {\bibfnamefont {J.~M.}\ \bibnamefont {Viana
  Parente~Lopes}}, \ and\ \bibinfo {author} {\bibfnamefont {N.~M.~R.}\
  \bibnamefont {Peres}},\ }\bibfield  {title} {\enquote {\bibinfo {title}
  {Gauge covariances and nonlinear optical responses},}\ }\href {\doibase
  10.1103/PhysRevB.96.035431} {\bibfield  {journal} {\bibinfo  {journal} {Phys.
  Rev. B}\ }\textbf {\bibinfo {volume} {96}},\ \bibinfo {pages} {035431}
  (\bibinfo {year} {2017})}\BibitemShut {NoStop}%
\bibitem [{\citenamefont {Vampa}\ \emph {et~al.}(2014)\citenamefont {Vampa},
  \citenamefont {McDonald}, \citenamefont {Orlando}, \citenamefont {Klug},
  \citenamefont {Corkum},\ and\ \citenamefont {Brabec}}]{Vampa2014}%
  \BibitemOpen
  \bibfield  {author} {\bibinfo {author} {\bibfnamefont {G.}~\bibnamefont
  {Vampa}}, \bibinfo {author} {\bibfnamefont {C.~R.}\ \bibnamefont {McDonald}},
  \bibinfo {author} {\bibfnamefont {G.}~\bibnamefont {Orlando}}, \bibinfo
  {author} {\bibfnamefont {D.~D.}\ \bibnamefont {Klug}}, \bibinfo {author}
  {\bibfnamefont {P.~B.}\ \bibnamefont {Corkum}}, \ and\ \bibinfo {author}
  {\bibfnamefont {T.}~\bibnamefont {Brabec}},\ }\bibfield  {title} {\enquote
  {\bibinfo {title} {Theoretical analysis of high-harmonic generation in
  solids},}\ }\href {\doibase 10.1103/PhysRevLett.113.073901} {\bibfield
  {journal} {\bibinfo  {journal} {Phys. Rev. Lett.}\ }\textbf {\bibinfo
  {volume} {113}},\ \bibinfo {pages} {073901} (\bibinfo {year}
  {2014})}\BibitemShut {NoStop}%
\bibitem [{elk()}]{elk}%
  \BibitemOpen
  \href@noop {} {\enquote {\bibinfo {title} {{The Elk Code}},}\ }\bibinfo
  {howpublished} {\url{http://elk.sourceforge.net/}}\BibitemShut {NoStop}%
\bibitem [{\citenamefont {Taillefer}(2010)}]{Taillefer2010}%
  \BibitemOpen
  \bibfield  {author} {\bibinfo {author} {\bibfnamefont {L.}~\bibnamefont
  {Taillefer}},\ }\bibfield  {title} {\enquote {\bibinfo {title} {{Scattering
  and Pairing in Cuprate Superconductors}},}\ }\href {\doibase
  10.1146/ANNUREV-CONMATPHYS-070909-104117} {\bibfield  {journal} {\bibinfo
  {journal} {http://dx.doi.org/10.1146/annurev-conmatphys-070909-104117}\
  }\textbf {\bibinfo {volume} {1}},\ \bibinfo {pages} {51--70} (\bibinfo {year}
  {2010})}\BibitemShut {NoStop}%
\end{thebibliography}%

\end{document}